\documentclass[a4paper]{article}
\usepackage{a4wide}
\usepackage{xr-hyper}
\usepackage{graphicx,subfigure,hhline}
	\graphicspath{{./}{figures/}}
\usepackage[colorlinks=true,linkcolor=blue,citecolor=red]{hyperref}
\usepackage{xcolor}
\usepackage{amssymb,amsthm,amsmath,bbold}
\usepackage{paralist}
\usepackage{cases}
\usepackage{booktabs}
\usepackage{braket}
\usepackage[strict]{changepage}
\usepackage[small,justification=justified]{caption}
\usepackage{titlesec} 
\usepackage{multicol}
\usepackage{lscape}
\usepackage{multirow}
\usepackage{makecell}
\usepackage[ntheorem]{empheq}
\usepackage[inline]{enumitem}
\usepackage[ruled,vlined,linesnumbered,algo2e]{algorithm2e}

\DeclarePairedDelimiter\ave{\langle}{\rangle}

\newcommand{\bU}{{\bf U}}

\newcommand{\bxi}{\boldsymbol{\xi}}
\newcommand{\Ub}{{\bf U}}

\newcommand{\bv}{{\bf v}}
\newcommand{\bV}{{\bf V}}
\newcommand{\hbV}{{\bf \hat{V}}}
\newcommand{\hv}{\hat{\bf v}}

\newcommand{\x}{{\bf x}}
\newcommand{\X}{{\bf X}}
\newcommand{\y}{{\bf y}}
\newcommand{\Y}{{\bf Y}}

\newenvironment{sistem}%
{\left\lbrace\begin{array}{@{}l@{}}}%
{\end{array}\right.}

\begin{document}

\title{A non-local kinetic model for cell migration: a study of the interplay between contact guidance and steric hindrance}
\author{Martina Conte, Nadia Loy}
\date{\today}

\maketitle

% REQUIRED
\begin{abstract}
We propose a non-local model for contact guidance and steric hindrance depending on a single external cue, namely the extracellular matrix, that affects in a twofold way the polarization and speed of motion of the cells. We start from a microscopic description of the stochastic processes underlying the cell re-orientation mechanism related to the change of cell speed and direction. Then, we formally derive the corresponding kinetic model that implements exactly the prescribed microscopic dynamics and, from it, it is possible to deduce the macroscopic limit in the appropriate regime. 
Moreover, we test our model in several scenarios. In particular, we numerically investigate the minimal microscopic mechanisms that are necessary to reproduce cell dynamics by comparing the outcomes of our model with some experimental results related to breast cancer cell migration. This allows us to validate the proposed modeling approach and, also, to highlight its capability of predicting the qualitative cell behaviors in diverse heterogeneous microenvironments. 
\end{abstract}

% REQUIRED
\textbf{keywords}
Non-local model, Kinetic transport equations, Markovian processes, Extracellular matrix, Steric hindrance, Contact guidance

% REQUIRED
\textbf{MSCcodes}
35Q20, 60J05, 92B05, 92C17

\section{Introduction}
It is well established that cell migration, based on diverse migration modes, is essential for normal processes such as embryonic development, immune function, and tissue repair, as well as it plays a critical role in disease states, including cancer dissemination \cite{ProvenzanoSciRep2016, provenzano2006collagen, Provenzano2009TCB, provenzano2008collagen, RayIB2018,Ray2017BioJ}. The process of cell migration is greatly affected by the surrounding microenvironment that cells sense through their protrusions and to which they respond by adapting their dynamics. 
A prominent role in cell migration is played by the extracellular matrix (ECM), which refers to the fiber-like components present within all tissues and organs and provides physical scaffolding for the cellular constituents. 
One of the major ECM component is collagen, which represents up to 30$\%$ of the total protein mass of a multicellular animal and, in particular, type I collagen is the most abundant one in the human body \cite{charras2014nat}.

\paragraph{The influence of the ECM on cell migration}
There are several biophysical and biochemical factors of ECM, in particular of collagens, that influence cell migration. The ones having a major impact are related to confinement, rigidity, topology, and adhesion properties \cite{charras2014nat}. Each one of these features gives rise to certain cell migration responses and it is often difficult from the experimental point of view to decouple the different aspects in order to investigate the effect and role of each of them separately. However, understanding every single mechanism and its specific role in the overall dynamics is important for extrapolating in vitro analyses to in vivo situations. For instance, one of the most outstanding goals in the context of cancer spread and dissemination is the development of therapeutic strategies targeting specific mechanisms that play a crucial role in cancer cell invasion. 

\noindent Concerning the influence of ECM on cell migration, it is possible to identify some factors that influence the direction of the cells and others that affect their speed. Specifically, the alignment of the collagen fibers is shown to stimulate {\it contact guidance}, \cite{Friedl_Brocker.00, Friedl.04} which is the tendency of cells to migrate by crawling on the fibers and following the directions imposed by them. 
On the other hand, EMC porosity - i.e., the spacing between fibers - affects cell speed \cite{charras2014nat}. In particular, it can lead to physical limits if the pores are too narrow, thus, representing a steric obstacle to cell motion. This phenomenon is known as \textit{steric hindrance}. Conversely, if the spacing between the fibers is larger than the cell size, then the cell starts having difficulties in forming the adhesion contact necessary for its motility. As the pores' average size may be related to the ECM density $M$, the mean speed may be expressed as a function of $M$ \cite{Wolf_Friedl.13}. In particular, it is found that there is an optimal matrix density $M_{max}$ assuring the maximum possible speed and a threshold value $M_{th}$, which corresponds to a small pore size, that hampers the cell from moving in a certain direction \cite{Wolf_Friedl.13}. Specifically, it may be shown that the mean speed has a quadratic-like dependence on the ECM density.
\paragraph{The role of the ECM in breast cancer dissemination}
Cell-ECM interactions have particular importance in the development and dissemination of breast cancer cells. In fact, the \textit{stromal} matrix surrounding tumors may be highly linearized and this would enhance cancer invasiveness \cite{han2016oriented,provenzano2006collagen,RayIB2018,Ray2017BioJ}. Thus, the study of cell response to a locally aligned matrix is of utmost interest, as it could suggest therapeutic strategies to target stromal invasion. In particular, in \cite{provenzano2006collagen} the authors introduce the concept of tumor-associated collagen signatures (TACS) that are used to stage mammary carcinoma tumor progression levels. Collagen-dense breast tissue increases also the risk of breast carcinoma, although the relationship between collagen density and tumorigenesis is not well-understood \cite{provenzano2008collagen}. 

\noindent Concerning the role of steric hindrance in cancer, in \cite{Plou} the authors perform experiments showing that increasing matrix density leads to reduced mean squared displacements and cell speeds (both mean and effective velocity). A first attempt to replicate these experimental results has been done in \cite{Goncalves}, where the authors propose a microscopic model in which they impose an external drag force to mimic ECM influence on cell speed. They assume that cells undergo increasing difficulty when migrating in denser, and consequently more viscous, matrices. This allows them to recover the fact that higher matrix densities imply lower speeds and mean squared displacements. Then, they impose a cubic net locomotive force with some ad hoc coefficients to recover the appropriate values of the speeds. Finally, in \cite{Ristori}, the authors investigate through a model the role of contact guidance and steric hindrance when in presence of cyclic stress.
\bigskip

\noindent As a matter of fact, the interplay between contact guidance and steric hindrance plays a significant role in breast cancer progression and dissemination. Provenzano, in particular, highlights the prominent role of these two aspects in his works \cite{provenzano2006collagen, provenzano2008collagen}. Notwithstanding, a systematic study of the coupling of these two aspects has not been investigated well. This is mainly related to the difficulty of building experimental settings in which the two mechanisms can be studied together as purely superposing effects. This is due, for instance, to the fact that the alignment of the fibers alters the porosity at the microscopic scale \cite{Taufalele2019PLOSONE}.  For the above reasons, we want to introduce a mathematical model that takes into account at the same time contact guidance and the impact of the ECM porosity/density on the cell speed. This would allow us to perform in silico experiments combining these two mechanisms and make predictions on the possible way cells sort or combine the two processes. 
Existing models regarding cell migration on ECM with a particular focus on the role of confinement and the influence of the ECM porosity/density on the cell speed include individual-based models \cite{Scianna_Preziosi.13, Scianna_Preziosi.14, Scianna2021Axioms,Scianna_Preziosi.13.2}, kinetic models \cite{Loy_Preziosi2}, and mechanical models \cite{preziosi2011mechanics,Preziosi2020Proc}. On the other hand, contact guidance has also been successfully described at the mesoscopic level through kinetic equations in \cite{Dickinson,Hillen.05}, where the authors propose models that allow to take into account the variation of the microscopic velocities in response to a given ECM fiber network. Kinetic equations have been proved to be very successful in modeling cell migration \cite{Calvez2015KRM,Chalub_Markowich_Perthame_Schmeiser.04,Chauviere_Hillen_Preziosi.07, Chauviere_Hillen_Preziosi.08, Filbet_Perthame, Filbet, Hillen.05,loy2021EJAM,loy2019JMB,Loy_Preziosi2,loy2020KRM}.

\noindent Kinetic models, in general, are intrinsically multiscale models. They allow to start from the microscopic description of the dynamics, including measurable parameters of the migration mechanism, and to derive a mesoscopic model in which the cells' position and velocity, which follow the prescribed microscopic dynamics, are statistically described by a distribution function. The time evolution of this function is ruled by a kinetic transport equation with a turning (or scattering) operator that implements the defined microscopic dynamics. By introducing the moments of the distribution function, it is also possible to derive macroscopic formulations describing the overall dynamics of the system.  
A particular class of kinetic equations is the one implementing {\it velocity-jump processes} \cite{Stroock}. These are microscopic Markovian processes that prescribe a \textit{transition probability} $T$ of choosing a new velocity and a \textit{frequency of re-orientation}  $\eta$, being, thus, $\frac{1}{\eta}$ the mean run time spent running on a linear tract between two consecutive re-orientations. Such equations are popularly used to model the cell migration mode called {\it run and tumble}, consisting in alternating runs over straight lines and re-orientations and they allow do a complete statistics regarding migration quantities (mean squared
displacement, mean and effective speed etc..) \cite{Alt.88}. 

\noindent Therefore, in this note we shall propose a non-local %kinetic
model for contact guidance and steric hindrance. As done in \cite{loy2019JMB,Loy_Preziosi2} we consider different processes for the speed and polarization of the cell, but in the present work, there is a single external cue, the ECM. The ECM is described statistically in terms of its macroscopic density and statistical distribution of the fiber direction affecting, respectively, the speed and direction of the cells. The ECM is evaluated non-locally in the physical space. The non-locality is due to the fact that cells sense the ECM and, thus, the direction of the fibers by extending their protrusions, which are a great determinant in contact guidance \cite{Carey2018} and, in general, in the presence of strongly heterogeneous or anisotropic environments \cite{han2016oriented,provenzano2006collagen,Sapudom2016AHM}. 
In particular, we state a microscopic discrete in time stochastic process from which we derive formally the kinetic model. To do this, we shall take advantage of classical tools of kinetic theory, mostly used in the literature of multi-agents systems \cite{Chiarello2022unp,pareschi2013BOOK}, which has its roots in the classical kinetic theory for gas dynamics and describes the dynamics through microscopic \textit{interaction rules} and \textit{collision-like kinetic Boltzmann-Povzner equations} for non-local interactions. In particular, such equations allow for a large variety of well-consolidated analytical tools, such as derivation procedures, the quasi-invariant limit, and limit scaling procedures, and they are poorly known in the community of cell migration modeling. Thus, in Section \ref{Micro_model} we describe more accurately the microscopic dynamics through these interaction rules. Then, in Section \ref{Meso_model}, after establishing a parallelism with the most known velocity jump process, we formally derive the kinetic model that implements exactly the microscopic dynamics. Moreover, in Section \ref{Macro_model}, we briefly review some classical procedures for deriving macroscopic models in the appropriate regime on the basis of the observed experimental parameters. Finally, we test our model in several scenarios in Section \ref{Num_sim}. We observe its ability to replicate different experimental results presented in \cite{Plou,tien2020} and related to breast cancer cell migration as well as to qualitatively predict the cell behavior in response to particular heterogeneous microenvironments.

\section{Mathematical modeling}
Our aim is to describe cell migration by modeling the re-orientation mechanism at the microscopic level and by the means of formally derived kinetic equations.
Each cell will be then identified by its position $\x \in \Omega \subset \mathbb{R}^{d}$, speed $v \in [0,U]$, being $U$ the maximum speed a cell can achieve, and polarization direction $\hv \in \mathbb{S}^{d-1}$, so that $\bv=v\hv\in [0,U]\times \mathbb{S}^{d-1}$ is the microscopic velocity vector. The distribution density function $p=p(t,\x,v,\hv)$, with $t>0$, will describe the statistical distribution of the speeds and directions for cells at time $t$ and located in $\x$.  

\noindent Macroscopic quantities describing the cell population can be defined as the \textit{statistical moments} of the distribution $p$, i.e.,

\noindent - the cell number (or macroscopic) density, which is the expected mass in $(t,\x)$,
\begin{equation}\label{def_rho}
\rho(t,\x) = \int_0^U \int_{\mathbb{S}^{d-1}} p(t,\x,v,\hv) \,d\hv\,dv
\end{equation}

\noindent - the mean velocity of cell located in $\x$ at time $t$
\begin{equation}\label{mean.U}
\Ub(t,\x) = \dfrac{1}{\rho(t,\x)}\int_0^U \int_{\mathbb{S}^{d-1}} p(t,\x,v,\hv)\bv \,d\hv\,dv\,.
\end{equation}

\noindent We are interested in cell migration on the ECM and, in particular, in two mechanisms: \textit{contact guidance}, which concerns the choice of the direction and depends on the fibrous structure of the ECM, and \textit{steric hindrance}, which affects cell speed and is regulated by the density of the ECM itself. Therefore, we also introduce the distribution function of the ECM fibers  
 $m=m(\x,\hv), \hv \in \mathbb{S}^{d-1}$, describing the statistical distribution of the fibers identified by their direction $\hv\in \mathbb{S}^{d-1}$ in each point of the physical space $\x\in\Omega\subset\mathbb{R}^d$.  As we do not consider re-modeling, the distribution $m$ is stationary. The macroscopic
density of the ECM is defined at each point $\x\in\Omega$ by
\begin{equation}
M(\x)=\int_{\mathbb{S}^{d-1}} m(\x,\hv) \, d\hv.
\end{equation}
Hence, the distribution
\begin{equation}
q(\x,\hv):=\dfrac{m(\x,\hv)}{M(\x)}
\end{equation}
 is, for each $\x\in \Omega$, the probability density function describing the statistical distribution of the directions of the fibers at $\x \in \Omega$.
In particular, as the fibers are not polarized, we assume that $q$ (and therefore $m$) is even as a function of $\hv$, $\forall \x \in\Omega$, which implies that the average direction of the fibers vanishes
\[
\mathbb{E}_q=\int_{\mathbb{S}^{d-1}} q(\x,\hv)\hv \, d\hv =\boldsymbol{0}.
\]
Moreover, we can introduce the variance-covariance matrix of $q$
\[
\mathbb{D}_q=\int_{\mathbb{S}^{d-1}} q(\x,\hv)\hv\otimes\hv \, d\hv. 
\]
When $q$ is a regular probability distribution, the tensor $\mathbb{D}_q$ is symmetric and positive definite and, thus, it is
diagonalizable. Equal eigenvalues correspond to an isotropic distribution of ECM. Conversely, an anisotropic distribution is characterized by different eigenvalues with the leading eigenvector representing the direction of preferential orientation of ECM fibers. This allows to reproduce isotropic/anisotropic migration on a
non-polarized fiber network \cite{Hillen.05,Painter2008}.

\subsection{Microscopic model}\label{Micro_model}
The individual dynamics of a cell may be described at the microscopic level by the means of evolution equations for random variables taking into account the position $\X_t\in\Omega$, the speed $V_t \in [0,U]$ and direction $\hbV_t\in \mathbb{S}^{d-1}$ of the cell, whose joint distribution function is $p(t,\X_t=\x,V_t=v,\hbV_t=\hv)$ for each $t$ and $\x$.  As classically done in kinetic theory, the microscopic dynamics of cell speed and direction is described by the means of \textit{binary interactions}. In the present case, the ECM fibers are described by the random variables $\Y_t \in \Omega$ and $\hbV^m_t \in \mathbb{S}^{d-1}$, whose distribution function is $m(\Y_t=\y,\hbV^m_t=\hv^m)$.

\noindent In particular, the re-orientation mechanism related to the change of cell speed and direction may be described in terms of discrete in time stochastic processes for the random variables $V_t$ and $\hbV_t$ which, during a time interval $\Delta t$, may change or not according to whether a re-orientation happens or not. These dynamics may be implemented in a discrete in time random process as 
\begin{equation}\label{eq:micro_stoch}
\begin{sistem}
\hbV_{t+\Delta t}=(1-\Sigma)\hbV_t+\Sigma \hbV_t'\\[0.2cm] 
V_{t+\Delta t}=(1-\Sigma)V_t+\Sigma V_t'
\end{sistem}
\end{equation}
where $\Sigma$ is a Bernoulli random variable with parameter $\mu B(\Y-\X)\Delta t$, saying whether a re-orientation, during which a cell changes both its direction and speed of motion, happens ($\Sigma=1$) or not ($\Sigma=0$). The quantity $\mu$ is the interaction frequency with the ECM fibers while $B$ is the interaction kernel taking into account non-local interactions in the physical space. We remark that in order for $\Sigma$ to be well defined, we need ${\Delta t \le 1/(\mu B(\Y-\X))}$, which means that for a high interaction frequency $\mu$ there is a high probability of having a re-orientation during a given time interval $\Delta t$. As in \cite{Chiarello2022unp}, we assume that $B$ has a compact support and that $\Delta t \le \dfrac{1}{\mu \max B}$. We stress the fact that in this microscopic process a cell simultaneously changes both the direction and the speed in a re-orientation.
The random variables $V_t'$ and $\hbV_t'$ denote the new speed and direction after a re-orientation. 
Classically in kinetic theory, the microscopic dynamics are described through \textit{interaction rules}.
In general, such microscopic rules are written in the form
\begin{equation}\label{eq:micro_boltz}
V_t'=I(V_t,\hbV_t,\hbV_t^m)+\sqrt{D}\Theta\, \qquad \hbV_t'=\hat{I}(V_t,\hbV_t,\hbV_t^m)+\sqrt{\mathbb{\hat{D}}}\Xi
\end{equation}
where $I,\hat{I}$ describe the deterministic part, which may depend on both the pre-re-orientation speed $V_t$, direction $\hbV_t$, and on the orientation of the fiber $\hbV^m_t$. $D$ and $\hat{\mathbb{D}}$ are diffusion matrices, being $\sqrt{\mathbb{\hat{D}}}^T\sqrt{\mathbb{\hat{D}}}=\mathbb{\hat{D}}$, while $\Theta$ and $\Xi$ are white noises, i.e., $\ave{\Psi}=\ave{\Xi}=0$, $\ave{\Psi^2}=\ave{\Xi^2}=1$. Here and hereafter, $\ave{\cdot}$ denotes the average operator. As we assume no ECM re-modeling, we have that in a binary interaction the direction of the fiber does not change, i.e.,
$\hbV^m_{t+\Delta t}=\hbV^m_t$.
This approach, which is classical in kinetic theory in the literature of multi-agent systems, allows for detailed descriptions of the microscopic dynamics. \\

\noindent \textbf{Remark} Usually, in works where experimental and computer-based models are coupled (e.g. \cite{Goncalves,Plou}), the evolution of the cell velocity is described by considering the superposing effect of external forces $\textbf{F}^{ext}$, modeled to take into account the influence of the external environment. In particular, these locomotive forces are determined by estimation from the data. This approach allows to use open-source softwares such as Physicell \cite{PhysiCell}, which is a hybrid 3D cell simulator that combines the model of the cellular environment (the chemical cues, the ECM mechanical behavior, etc.) as a continuum with  an agent-based model for the cells. The approach we propose here allows to start from an agent-based model in which details about the microscopic dynamics may be implemented and, then, included in the macroscopic models that will be derived. In the modeling framework given by \eqref{eq:micro_boltz}, $I$ and $\hat{I}$ may be linked to the external forces acting on the cells by simply setting $V_{t+\Delta t}\hbV_{t+\Delta t}=V_t\hbV_t +\textbf{F}^{ext} \Delta t$.\\

\noindent If we want to implement a velocity-jump process, we need to consider transition probabilities as probability density functions of the random variables $V_t'$ and $\hbV_t'$ that are given by
\begin{equation}\label{eq:micro_trans}
V_t'\sim \psi(V_t'|M(\Y_t)), \qquad \hbV_t' =\hbV^m_t\sim m(\Y_t,\hbV^m_t)\,.
\end{equation}
Here, $\psi=\psi(v|M)$ is a probability density function of the speeds, conditioned by the ECM density $M$, and it has an average speed $\bar{v}_M$ depending on $M$ and a second moment $D_M$ such that its variance is $D_M-\bar{v}_M^2$.

\noindent We then consider the kinematic relation for the variation of the position $\X_t$ during a time interval $\Delta t$ given by
\begin{equation}\label{eq:space_ev}
\X_{t+\Delta t}=\X_{t}+\Delta t\bV_{t}
\end{equation}

 \noindent As already mentioned, the ECM is explored by the cell through its protrusions, which may be  extended up to a maximum \textit{sensing radius} $R$. As previously done in \cite{Loy_Preziosi2}, in order to include physical limits of migration, we shall consider a non-constant sensing radius, identifying the fact that a cell cannot measure the external cue in a physical region that cannot be reached. In particular, the ECM density $M_{th}$ represents this physical limit and we shall consider the non-constant sensing radius defined by \cite{Loy_Preziosi2}
\begin{equation}\label{def:PLM}
R_{M}(t,\x,\hv)=
\begin{sistem}
R\quad\quad \text{if}\quad M(t,\x+\lambda\hv)\le M_{th}\quad \forall \lambda \in [0,R],\vspace{0.7cm}\\
\inf\{\lambda\in[0,R]\,:\,M(t,\x+\lambda\hv)>M_{th}\}\quad \text{otherwise.}
\end{sistem}
\end{equation}
The latter means that, in a given direction $\hv$, the sensing radius is limited as soon as the protrusion encounters a region that the cell cannot migrate through, while it is maximum if such a region, in a certain direction $\hv$, is not reached within a distance $R$.
In particular, the sensing radius will affect the support of the interaction kernel, also called Povzner kernel, namely
\begin{equation}\label{def:Povzner}
B(\y-\x)=\delta(\y-(\x+\lambda \hv)) \gamma(\lambda)
\end{equation}
where $\gamma(\lambda)$ is a sensitivity function having compact support in $[0,R_M(t,\x,\hv)]$ that weights the information given by $m$ in each direction $\hv$. In particular it may be a Dirac delta if the cell only evaluates the information on the tip of the protrusion, while it may be a characteristic function if the cell weights uniformly the information up to the tip of the protrusion. 

\subsection{Mesoscopic model}\label{Meso_model}
Through a rather classical procedure \cite{pareschi2013BOOK}, it is possible to derive a kinetic equation for the evolution of the distribution $p$ describing the statistical distribution of cells obeying to the microscopic process \eqref{eq:micro_stoch}-\eqref{eq:space_ev}-\eqref{def:Povzner} joined with \eqref{eq:micro_boltz} or \eqref{eq:micro_trans}.

\noindent Let $\phi=\phi(\x,v,\hv)$ be an observable quantity defined on $\Omega\times [0,U]\times \mathbb{S}^{d-1}$ that we consider to be a $\mathcal{C}^{\infty}$ function having compact support. From~\eqref{eq:micro_stoch} together with the assumed independence of $\Sigma$, we see that the mean variation rate of $\phi$ in the time interval $\Delta{t}$ satisfies
\begin{align*}
	&\frac{\ave{\phi\left(\X_{t+\Delta t},V_{t+\Delta t}, \hbV_{t+\Delta t}\right)}-\ave{\phi\left(\X_t,V_{t},\hbV_t\right)}}{\Delta{t}}= \\
	&\frac{\ave{\phi\left(\X_{t}+V_t\hbV_t\Delta t,(1-\Sigma)V_{t} + \Sigma V_t',(1-\Sigma)\hbV_{t} + \Sigma \hbV_t'\right)}-\ave{\phi\left(\X_t,V_t,\hbV_t\right)}}{\Delta{t}}=\\
	&\frac{\ave{\phi\left(\X_{t}\!\!+\!\!V_t\hbV_t\Delta t,V_{t},\hbV_t\right)(1\!\!-\!\!\mu B \Delta t)\!+\!\phi\left(\X_t\!\!+\!\!V_t\hbV_t\Delta t,V_t',\hbV_t'\right)\mu B \Delta t\!\!-\!\!\phi\left(\X_t,V_t,\hbV_t\right)}}{\Delta{t}},
\end{align*}
whence we deduce the instantaneous time variation of the average of $\phi$ in the limit $\Delta{t}\to 0^+$ as
\begin{equation}\label{eq:ave_1} 
\begin{aligned}[b]
\frac{d}{dt}\ave{\phi\left(\X_t,V_t,\hbV_t\right)}=&\ave{\mu B(\Y_t-\X_t) \left[\phi\left(\X_t,V_t',\hbV_t'\right)-\phi\left(\X_t,V_t,\hbV_t\right)\right]}\\
&-\nabla_\x \cdot \ave{V_t\hbV_t \phi(\X_t,V_t,\hbV_t)}\,.
\end{aligned}
\end{equation}
Here, on the right hand side the first two terms take into account the gain and loss terms related to the re-orientation process, respectively, while the last term accounts for the free particle drift.
 If the microscopic process ruling the evolution of $V_t'$, $\hbV_t'$ is a velocity jump process \eqref{eq:micro_trans}, then the gain term describing the new cell acquired speed and direction $(V_t',\hbV_t')$ can be written as
\begin{equation}\label{eq:1}
\begin{aligned}[b]
&\ave{B(\Y_t-\X_t)\phi\left(\X_t,V_t',\hbV_t'\right)}=\\
&=\ave{\int_{\Omega^2}\!\int_{0}^U\!\int_{{\mathbb{S}^{d-1}}^2}\! B(\y-\x) \phi(\x,v',\hv')\psi(v'|M(\y)) m(\y,\boldsymbol{n})
 p(t,\x,v,\hv) \, d\boldsymbol{n} d\hv dv d\x d\y }\\
 &=\ave{\int_{\mathbb{R}_+}\!\!\int_\Omega \!\int_{\mathbb{S}^{d-1}\times[0,U]}\!\!\!\!\!\!\!\!\!\!\!\!\!\!\!\!\!\!\!\phi(\x,v',\hv')\psi(v'|M(\x+\lambda \hv')) m(\x+\lambda \hv',\hv')\gamma(\lambda) \,  p(t,\x,v,\hv)
d\hv dv d\x d\lambda }, 
 \end{aligned}
\end{equation}
while 
\begin{equation}\label{eq:2}
\begin{aligned}[b]
&\ave{B(\Y_t-\X_t)\phi\left(\X_t,V_t,\hbV_t\right)}=\\
&=\int_\Omega\int_\Omega\int_{0}^U B(\y-\x) \int_{\mathbb{S}^{d-1}} \phi(\x,v,\hv)\, p(t,\x,v,\hv) m(\y,\hv) \, d\hv  dv d\x d\y, 
\end{aligned}
\end{equation}
and 
\begin{equation}\label{eq:3}
\nabla_\x \cdot \ave{V_t\hbV_t \phi(\X_t,V_t,\hbV_t)}=\nabla_\x\cdot \int_\Omega\int_{0}^U \int_{\mathbb{S}^{d-1}} v\hv \phi(\x,v,\hv)\, p(t,\x,v,\hv) \, d\hv dv   d\x. 
\end{equation}
Rewriting \eqref{eq:ave_1} with \eqref{eq:1}-\eqref{eq:2}-\eqref{eq:3} and choosing  $\phi(\x,v,\hv)=\xi(\x) \varphi(v,\hv)$, the kinetic equation for $p$ is
\begin{equation}\label{eq:transp_weak}
\begin{aligned}[b]
&\dfrac{d}{dt}\int_{0}^U \int_{\mathbb{S}^{d-1}}\varphi(v,\hv) p(t,\x,v,\hv) d\hv dv+\nabla_\x \cdot \int_{0}^U \int_{\mathbb{S}^{d-1}} \varphi(v,\hv) \bv p(t,\x,v,\hv) \, d\hv dv=\\
&\eta\ave{\int_{0}^U \int_{\mathbb{S}^{d-1}}T[m](v',\hv')\varphi(v',\hv')\, d\hv' dv'-\int_{0}^U \int_{\mathbb{S}^{d-1}}\varphi(v,\hv)p(t,\x,v,\hv) \,d\hv  dv }
\end{aligned}
\end{equation}
where
\begin{equation}\label{eq:trans1}
T[m](\x,v,\hv)= \int_{0}^{R_M(\x,\hv)} \dfrac{m(\x+\lambda \hv,\hv)}{\bar{M}(\x)}\psi(v|M(\x+\lambda \hv)) \gamma(\lambda) \, d\lambda
\end{equation}
is a \textit{transition probability} satisfying
\[
\int_0^U \int_{\mathbb{S}^{d-1}}T[m](\x,v,\hv) \,d\hv\,dv =1
\]
and describing the probability for a cell located at $\x$ of choosing a speed $v$ and direction $\hv$.
Such a transition probability encodes the fact that a cell extends its protrusions and senses the ECM in each direction collecting at the same time information regarding the fibers structure and macroscopic ECM density and weights them in the same way.  The strong form of equation \eqref{eq:transp_weak} is
\begin{equation}\label{eq:transp}
\dfrac{\partial p}{\partial t} (t,\x,v,\hv)+\bv \cdot \nabla p (t,\x,v,\hv)=\mathcal{J}[p](t,\x,v,\hv)
\end{equation}
that describes the evolution of the statistical distribution of the cells that obey the microscopic dynamics \eqref{eq:micro_trans}, where the right hand side is the \textit{turning operator}
\begin{equation}\label{eq:turning_operator}
\mathcal{J}[p](t,\x,v,\hv)=\eta(\rho(t,\x)T[m](\x,v,\hv)-p(t,\x,v,\hv))\,.
\end{equation}
In \eqref{eq:turning_operator}, $\eta=\mu\bar M$ is the \textit{frequency of re-orientation}, which depends not only on the interaction frequency but also on the measured quantity of ECM, thus showing that the microscopic process \eqref{eq:micro_stoch}-\eqref{eq:micro_trans} implies a re-orientation mechanism happening with a frequency that depends on the sensed (on the whole neighborhood) ECM density.
The quantity
\[
\bar{M}(\x)=\int_{\mathbb{S}^{d-1}}\int_0^{R_M(\x,\hv)} m(\x+\lambda \hv,\hv) \gamma(\lambda) d\lambda d\hv
\]
is, in fact, an average of the density of the ECM over the measured neighborhood where the information is weighted by $\gamma$. We stress the fact that this particular transition probability \eqref{eq:trans1} is the one implementing the microscopic process \eqref{eq:micro_stoch}-\eqref{eq:micro_trans}, in which cells change simultaneously both the direction and speed as they sense the same external cue $m$, that influences through two different mechanisms the choice of the direction and the speed. This is different from considering two independent sensings as in \cite{loy2019JMB,Loy_Preziosi2}, where the two measured quantities affecting the direction and the speed have different origins. The average velocity of the transition probability \eqref{eq:trans1} is given by
\begin{equation*}
\begin{aligned}[b]
\bU_T(\x)&= \int_{\mathbb{S}^{d-1}}\int_0^UT[m](\x,v,\hv)\,v \hv\,dvd\hv\\
&=\int_{\mathbb{S}^{d-1}} \int_{0}^{R_M(\x,\hv)}\dfrac{m(\x+\lambda \hv,\hv)}{\bar{M}(\x)} \bar{v}(\x|M(\x+\lambda \hv)) \gamma(\lambda) \, d\lambda \, \hv \, d\hv 
\end{aligned}
\end{equation*}
while its variance-covariance matrix is
\begin{equation*}
\begin{split}
\mathbb{D}_{T}(\x)&= \int_{\mathbb{S}^{d-1}}\int_0^U T[m](\x,v,\hv)\,(\bv-\bU_T)\otimes (\bv-\bU_T)\,dv\,d\hv =\\[0.3cm]
&=D_M\int_{\mathbb{S}^{d-1}} \int_0^{R_M(\x,\hv)}\dfrac{m(\x+\lambda \hv,\hv)}{\bar{M}(\x)}\gamma(\lambda) \, d\lambda\,\hv\otimes \hv\,d\hv-\bU_{T}(\x)\otimes \bU_{T}(\x)
\end{split}
\end{equation*}
where $D_M$ is the energy of the probability density function $\psi$ of the speeds. We assume that it is constant, as all the cells are affected with the same degree of stochastic variation. \\%

\noindent \textbf{Remark} 
If we want to implement the dynamics \eqref{eq:micro_boltz}, then the gain term is 
\[
\begin{aligned}[b]
&\ave{B(\Y_t-\X_t)\phi\left(\X_t,V_t',\hbV_t'\right)}=\\&=\ave{\int_{\Omega^2}\int_{0}^U\int_{{\mathbb{S}^{d-1}}^2}  B(\y-\x) \varphi(\x,v',\hv')\gamma(\lambda)p(t,\x,v,\hv) m(\y,\boldsymbol{n}) \, 
d\boldsymbol{n}  d\hv dv d\x d\y}
\end{aligned}
 \]
with $v', \hv'$ given by \eqref{eq:micro_boltz}, so that the kinetic equation is the Boltzmann-Povzner equation \cite{Povzner,Fornasier2011PhD}
\begin{equation}\label{eq:Boltz_coll}
\begin{aligned}[b]
&\dfrac{d}{dt}\int_{0}^U \int_{\mathbb{S}^{d-1}}\varphi(v,\hv) p(t,\x,v,\hv) dv d\hv+\nabla_\x \cdot \int_{0}^U \int_{\mathbb{S}^{d-1}} \varphi(v,\hv) \bv p(t,\x,v,\hv) \, dv d\hv=\\
&\eta\ave{\int_{0}^U \int_{\mathbb{S}^{d-1}}\varphi(v',\hv')-\varphi(v,\hv)p(t,\x,v,\hv) \, dv d\hv}
\end{aligned}
\end{equation}
In the case $B=1$,
choosing
\begin{equation}\label{eq:comp}
I(v,\hv,\hv_n)=\bar{v}, \ D=D_M-\bar{v}^2, \quad \hat{I}(v,\hv,\hv_n)=\mathbb{E}_q,  \ \hat{\mathbb{D}}=\mathbb{D}_q^T\mathbb{D}_q
\end{equation}
in \eqref{eq:micro_boltz}, then the average and energy of $p$ prescribed by the model \eqref{eq:Boltz_coll}-\eqref{eq:micro_boltz} is the same as
 if its evolution were ruled by \eqref{eq:transp}-\eqref{eq:turning_operator}-\eqref{eq:trans1}. It is worth mentioning that the average and energy of $p$ are the ones that are involved in the hydrodynamic description of the system. Moreover, the microscopic model \eqref{eq:micro_stoch}-\eqref{eq:micro_boltz}-\eqref{eq:space_ev}-\eqref{def:Povzner} with the choice \eqref{eq:comp} instead of \eqref{eq:micro_trans} may be particularly useful in the case in which $q$ and $\psi$ are not easy to sample.

\subsection{Macroscopic equations}\label{Macro_model} 
In order to investigate the overall trend of the system, the macroscopic behavior is typically analyzed. 
This is done through the derivation of macroscopic evolution equations for $\rho(t,\x)$ in the emerging regime of the system, which may result from a proper non-dimensionalization of the system.  
Formally, we introduce a small parameter $\epsilon \ll 1$ and we re-scale the spatial variable as 
\begin{equation}\label{eq:scale_space}
\boldsymbol{\xi}=\epsilon \x,
\end{equation} 
being $\boldsymbol{\xi}$ the macroscopic spatial variable.
According to the other characteristic quantities of the system and up to an appropriate non-dimensionalization, the macroscopic time scale $\tau$ will be 
\begin{equation}\label{eq:diff_scale}
\tau=\epsilon^{3-\gamma} t,
\end{equation}
The appropriate scaling of the system can be done by analyzing its dominant behavior, which can be investigated by measuring the mean squared displacement (MSD)
\begin{equation}\label{def:MSD}
MSD(t):=\ave{||\x||^2}
\end{equation}
and determining its growth with respect to time, i.e.,
\begin{equation}
\ave{||\x||^2} \sim t^\gamma
\end{equation}
where
\begin{itemize}
\item $\gamma=1$ indicates a diffusion dominated phenomenon (purely diffusive);
\item $\gamma=2$ indicates a drift dominated phenomenon (purely directed). 
\end{itemize}
The two choices correspond  to a parabolic scaling ($\tau=\epsilon^2t$) and to a hyperbolic scaling ($\tau=\epsilon t$), respectively. Therefore, we shall consider a diffusive or a hydrodynamic scaling of the transport equation \eqref{eq:transp} with \eqref{eq:turning_operator}.  These limits techniques relying on Hilbert expansions for transport equations with velocity jump processes have been widely treated in \cite{Hillen.05,Othmer_Hillen.00, loy2019JMB, Loy_Preziosi2,Othmer_Hillen.02}. They are based on expansions of the transition probability as
\[
T(\boldsymbol{\xi},v,\hv)=T_0(\boldsymbol{\xi},v,\hv)+\epsilon T_1(\boldsymbol{\xi},v,\hv)+\mathcal{O}(\epsilon^2).
\]
and, consequently, of its average $\Ub_T^i$, variance-covariance matrix $\mathbb{D}_T^i$, and of the distribution function $p$. In particular, the fundamental property for performing the diffusive limit requires that the leading order of the drift vanishes, i.e., 
\begin{equation}\label{UT0.0}
\Ub^0_T=0.
\end{equation}
Carrying out the asymptotic procedure leads to
\begin{equation}\label{macro_diff}
\dfrac{\partial}{\partial {\tau}} \rho +\nabla \cdot \left( \Ub_T^1 \rho\right)=\nabla \cdot \left[ \dfrac{1}{\eta} \nabla \cdot \left(\mathbb{D}_T^0 \rho\right) \right]\,,
\end{equation}
being $\mathbb{D}_T^0$ 
the diffusion motility tensor. %and recalling \eqref{rho0}.
Equation \eqref{macro_diff} is a diffusion-advection equation, where $\Ub_T^1$ is the drift velocity of first order.

\noindent If \eqref{UT0.0} does not hold, as typically happens if $R$ is large with respect to the length of variation of the external field $m$, but the non-dimensionalization of the system or experimental observations prescribe a diffusive regime,  we can consider a drift-diffusion limit as it was done in \cite{loy2021EJAM}. Setting $p(\tau,\bxi,\bv)=u(\tau,z,\bv)$, with $z=\bxi-\bU_T\tau$, we have
\[
 \dfrac{\partial}{\partial {\tau}} p+\bv\cdot\nabla p=\mathcal{L}[p]\,\,\Longleftrightarrow\,\,\dfrac{\partial}{\partial {\tau}}  u+\nabla\cdot((\bv-\bU_T)u)=\mathcal{L}[u]
\]
Going back to the original variable $p$ and remembering \eqref{rho0}, we get
\begin{equation}\label{macro_diffdrift}
\dfrac{\partial}{\partial {\tau}} \rho+\nabla\cdot(\bU_T\rho)=\nabla\cdot\left(\dfrac{1}{\eta}\nabla\cdot(\mathbb{D}_T\rho)\right)
\end{equation}

\noindent If, instead, a hyperbolic scaling is required, we can use the results presented in \cite{Hillen.05} that gives
\begin{equation}\label{macro_hyp}
 \dfrac{\partial}{\partial {\tau}} \rho + \nabla\cdot (\rho \bU_T)=\varepsilon\nabla\cdot\left(\dfrac{1}{\eta}\nabla\cdot(\mathbb{D}_T\rho)+\dfrac{1}{\eta}\rho\,\bU_T\nabla\cdot\bU_T\right)\,.
\end{equation}
This is the equation with the first-order correction in which we can recover the dependency on the ECM through the frequency $\eta$ in the correction term.

\noindent Major details about the well-known techniques required for the asymptotic procedures are reported in the Supplementary Material \ref{Macro_model_SM} for the reader's convenience.

\section{Numerical investigations}\label{Num_sim}

In this section we present some numerical tests. In particular, we shall integrate numerically 
\begin{itemize}
\item the microscopic model \eqref{eq:micro_stoch}-\eqref{eq:space_ev}-\eqref{def:Povzner} with Monte Carlo methods as in \cite{loy2021KRM}; 
\item the kinetic model \eqref{eq:transp}-\eqref{eq:turning_operator}-\eqref{eq:trans1} with the same method used in \cite{loy_conte2020,loy2019JMB,Loy_Preziosi2};
\item the macroscopic diffusion and drift-diffusion models \eqref{macro_diff} or \eqref{macro_diffdrift}  with a continuous Galerkin finite element scheme \cite{quarteroni2008numerical}, while the drift model \eqref{macro_hyp} with a Donor-cell advection scheme \cite{Leveque}.
\end{itemize}
Concerning the boundary conditions \cite{Plaza}, we shall consider no flux boundary conditions, that for $p$, may be given, for example, by specularly reflective boundary conditions.

\noindent We shall present four numerical tests:
\begin{itemize}
\item[\textbf{Test 1}] in Section \ref{sec.num.1} we validate the microscopic model \eqref{eq:micro_stoch}-\eqref{eq:space_ev}-\eqref{def:Povzner} with the choice \eqref{eq:micro_trans} (with $B=1$ as we are on a spatially homogeneous setting) by comparing simulations with the experimental results presented in \cite{Plou}, where the authors investigate the phenomenon of steric hindrance on collagen gel. We remark that, as we are in a homogeneous setting, the microscopic model \eqref{eq:micro_stoch}-\eqref{eq:space_ev} with the choice \eqref{eq:micro_boltz} forecasts the same mass and average velocity for a large number of particles and a small $\Delta t$;
\item[\textbf{Test 2}] in Section \ref{sec.num.2} we consider an application of our microscopic model to the invasion of breast cancer cells from an aggregate into the collagen according to the experiments presented in \cite{tien2020} and we compare the results of the microscopic and kinetic models. We also provide comparisons with the corresponding macroscopic limits;
\item[\textbf{Test 3}] in Section \ref{sec.num.3} we apply our model in order to investigate and make predictions on the dynamics of cells moving on collagen fibers with different densities and fiber alignment; 
\end{itemize}
Moreover, in the Supplementary Section \ref{sec.num.4}, we include a further test ({\bf Test 4}) in order to investigate with the kinetic model a heterogeneous environment with an interface dividing regions with different collagen densities and/or fiber alignment. This is a more qualitative analysis that shows the potential applicability of our approach.
\subsection{Steric hindrance on collagen gel}\label{sec.num.1}
\begin{figure}[!h]
\begin{center}
\subfigure[Case $i)$]{\includegraphics[width=0.3\textwidth]{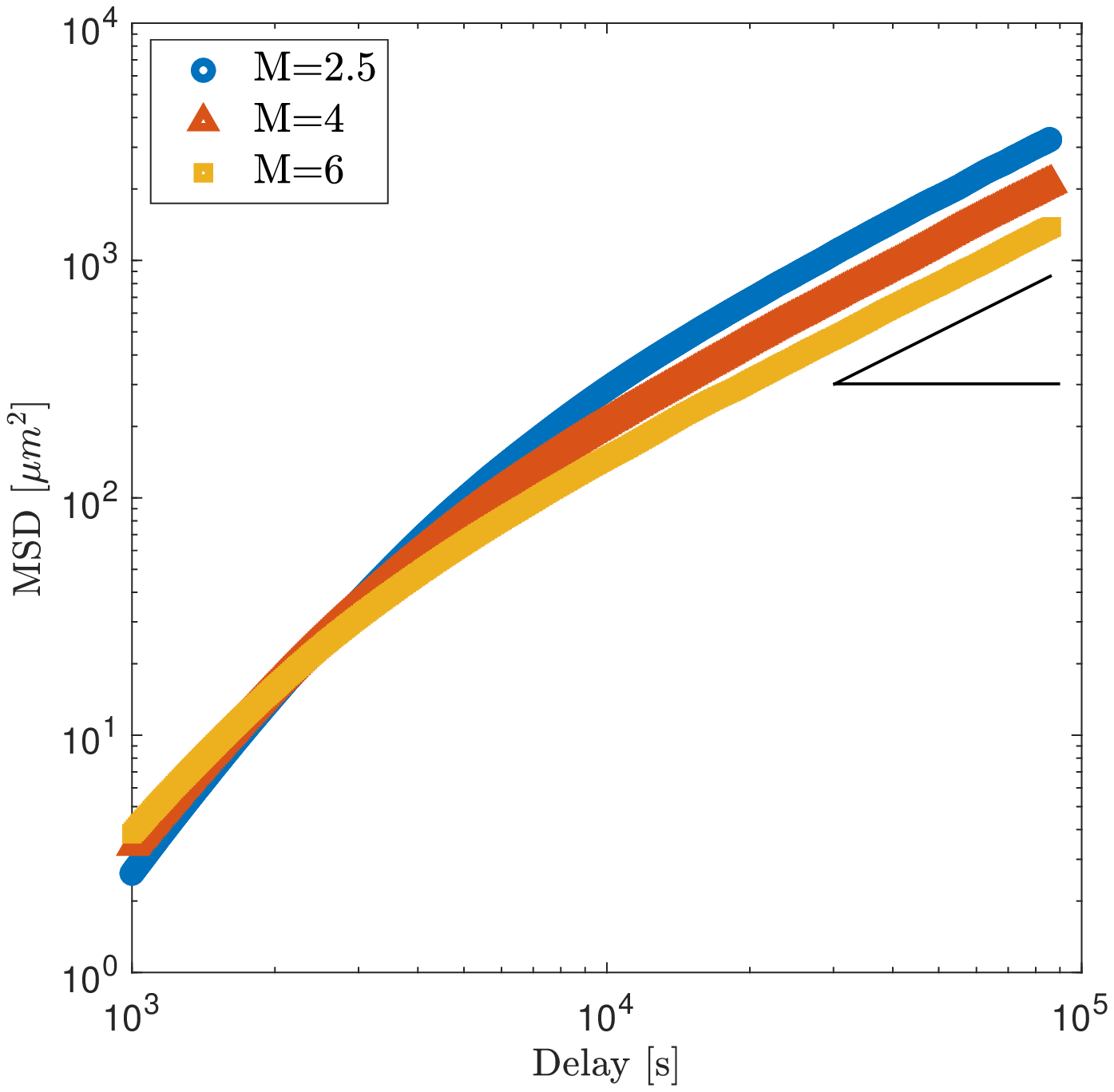}}
\subfigure[Case $ii)$]{\includegraphics[width=0.3\textwidth]{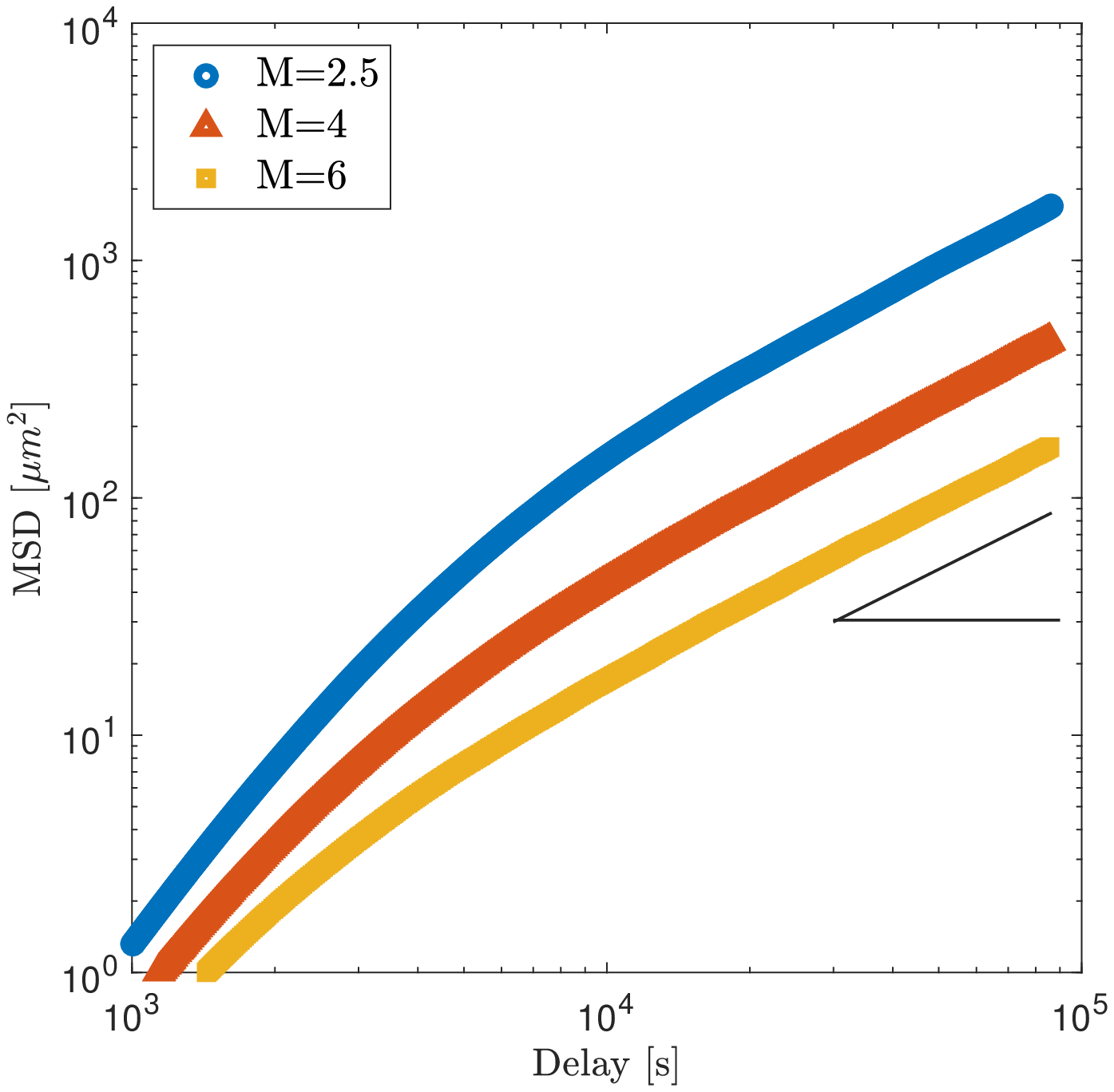}}
\subfigure[Case $iii)$]{\includegraphics[width=0.3\textwidth]{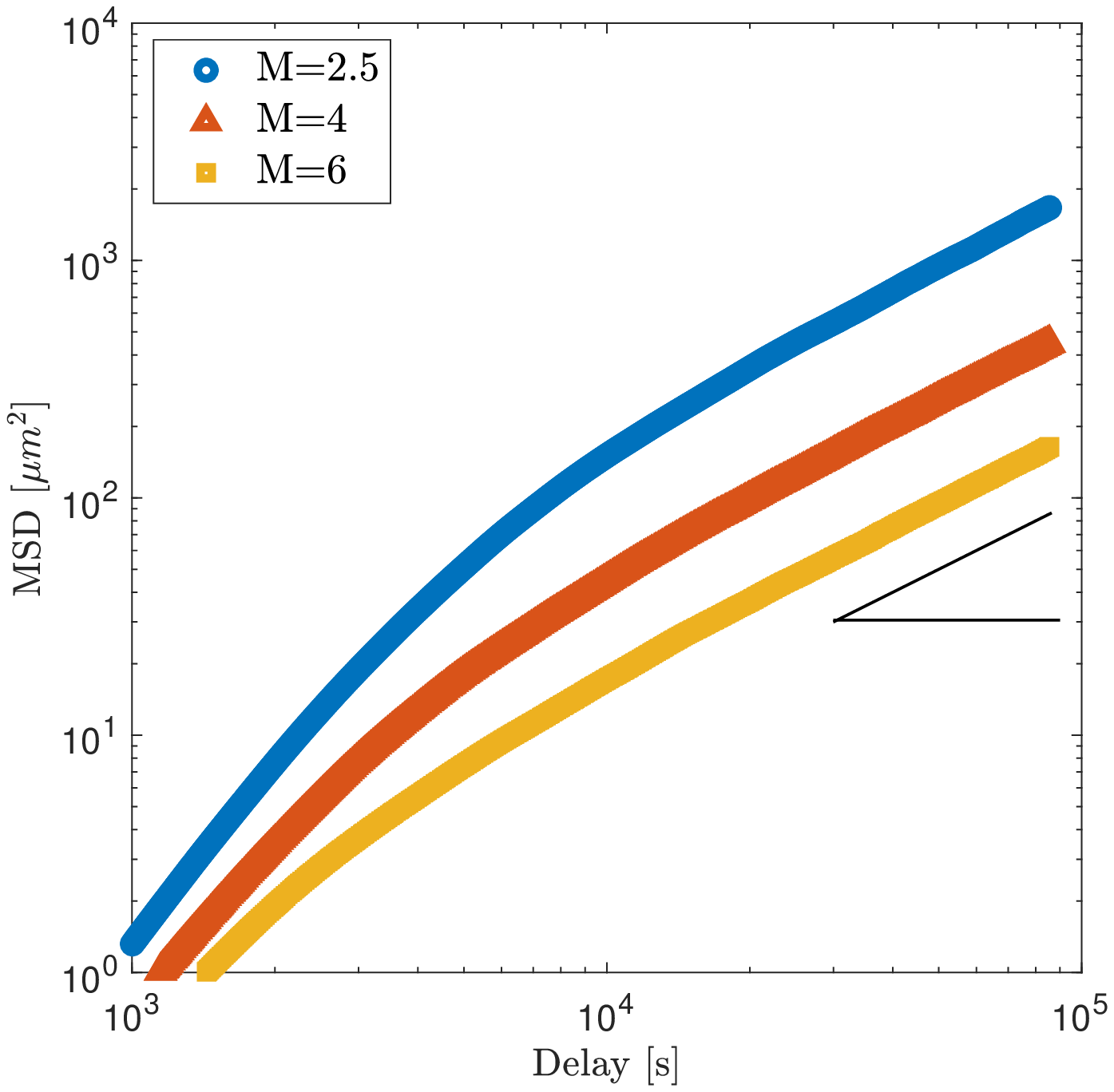}}
\vspace{-0.3cm}
\caption{Mean squared displacement for the macroscopic densities $M=[2.5,\,4,\,6] \,mg/mL$ in the following cases: $i)$ uniform speed and random fibers in (a); $ii)$ $M$-dependent speed and random fibers in (b); $iii)$ $M$-dependent speed and aligned fibers in (c). Simulations are run for $24$ h with $\Delta t=0.0167$ h.}
\label{fig_msd_SR}
\end{center}
\end{figure}

Firstly, we consider the experimental results presented in \cite{Plou}. Here, the authors track every $20$ minutes for $24$ hrs 50 NSCLC (Non-Small Cell Lung Cancer) cells moving on a 3D collagen-based matrix,  made up of a collagen type I from bovine skin medium of different densities. In particular, time-lapse images are acquired from the focal plane located in the middle of the $z$-axis, while out-of-focus cells are not quantified. Thus, the performed analysis on cell motility is substantially quantified in a 2D scenario. These experiments show how collagen density affects the strength of the physical barrier. Precisely, it interferes with cell migration by trapping single metastatic NSCLC cells and preventing their dissemination through the matrix. The authors find that for increasing values of the ECM density, the cell mean speed decreases and, correspondingly, the mean squared displacement becomes lower. They consider fixed collagen concentrations of 2.5 $mg/mL$, 4 $mg/mL$, and 6 $mg/mL$ and measure the corresponding average speeds, given by $\bar{v}_M=[0.1696,0.104,0.063] \,\mu m/ min$. We use the presented modeling framework to replicate these experiments, looking at the minimal combination of ingredients that would allow to obtain comparable results.\\
\noindent To this aim, we analyze three different settings considering
\begin{itemize}
\item[$i)$] an $M$-dependent frequency for the cell turning, a uniform speed distribution, and a random fiber network;
\item[$ii)$] an $M$-dependent frequency and speed distribution, combined with a random fiber network;
\item[$iii)$] an $M$-dependent frequency and speed distribution as well as an aligned fiber network.
\end{itemize} 
Precisely, the density dependent frequency is given by $\eta=\mu M$, as recovered in the derivation of \eqref{eq:turning_operator}. For the uniform speed distribution over $[0,U]$ in \eqref{eq:micro_trans}, we consider 
\begin{equation}
\psi(v):=\frac{1}{U}
\label{unif_speed_distr}
\end{equation} 
(corresponding to $\bar{v}_M=U/2, D_M=U^2/12$). Instead, for the case of density-dependent speed distribution we analyze two possible choices for $\psi(v|M)$: a unimodal von Mises distribution rescaled over $[0,U]$
\begin{equation}
\psi(v|M)=\frac{1}{2\pi I_0(k)} \exp\left[k_\psi \cos\left(2\pi\frac{v-\bar{v}_M}{U}\right)\right]\,
\label{uvm}
\end{equation}
where $k_\psi$ is the concentration parameter, $I_0(k)$ the Bessel function of order 0, and where we impose the value of the mean speed $\bar{v}_M$ for the different values of the matrix density; a truncated Gaussian distribution, defined on $[0,U]$
\begin{equation}
\psi(v|M)=
\begin{sistem}
\dfrac{1}{\sigma}\dfrac{\phi\left(\dfrac{v-\nu}{\sigma}\right)}{\Phi\left(\dfrac{U-\nu}{\sigma}\right)-\Phi\left(\dfrac{-\nu}{\sigma}\right)}\qquad \text{for}\,\,0\le v\le U\\[1cm]
0\hspace{4.6cm} \text{otherwise}\,,
\end{sistem}
\label{tnd}
\end{equation}
where $\phi$ is the probability density function of the standard normal distribution, $\Phi$ its cumulative distribution function, and we impose the values of $\sigma$ and $\nu$, parameters related to the mean and variance of the distribution\footnote{If we consider a random variable $\mathcal{X}$ with normal distribution with mean $\nu$ and variance $\sigma^2$ and lying within the interval $-\infty\le0<U\le\infty$, then $\mathcal{X}$ conditional on $0<\mathcal{X}<U$ has a truncated normal distribution with parameter $\nu$ and $\sigma$.}. Finally, for the fiber network, we describe a random fiber distribution by assuming
\begin{equation}
q(\theta)=\dfrac{1}{2\pi}\,,
\label{rand_fiber}
\end{equation}
while for the aligned fibers we use a bimodal von-Mises distribution, with given concentration parameter $k>0$ and preferential direction of migration $\theta_q$
\begin{equation} %\bvm (w) 
q(\theta)= \frac{1}{4\pi I_0(k)} \Big(\exp\left[k \cos(\theta-\theta_q)\right] + \exp\left[-k \cos(\theta- \theta_q)\right] \Big).
\label{bvm} 
\end{equation}
For each case, we investigate both the mean square displacement (MSD) of the cells and the cell tracking, considering the same experimental settings proposed in \cite{Plou}. Moreover, we evaluate the variation of cell mean and effective speed in relation to the ECM density. Precisely, we consider a domain $\Omega=[-150,150]\times[-150,150] \,\mu m^2$ with $10^{4}$ cells moving with maximum speed $U=0.4\, (\mu m/ min)$ and $\mu=1.8\cdot 10^{-2} \, (1/min)$, which corresponds to a frequency of about 1 ($1/hr$). Results about the MSD evolution, obtained with the integration of \eqref{eq:micro_stoch}-\eqref{eq:space_ev}-\eqref{def:Povzner} with \eqref{eq:micro_trans} and the aforementioned $q$ and $\psi$, are shown in Fig. \ref{fig_msd_SR}.

\noindent In all cases, we observe that an increase in the matrix density leads to a decrease in the cell mean squared displacement, meaning that including an $M$-dependent frequency is sufficient to recover this feature. However, to obtain more accurate results in terms of final MSD value and differences among the three values of $M$ density (as reported in \cite{Plou}), including an $M$-dependent speed distribution appears to be fundamental. This is also necessary to get the different behavior of the mean speed reported in the experiments. In the first case $i)$, in fact, cells always have an average speed given by $U/2$. In the case $ii)$, instead, using \eqref{uvm} we are able to recover the appropriate effective speed in the different cases, as reported in the Supplementary Table \ref{Veff_VM}.  Moreover, setting $\nu:=U/M$ and  $\sigma:=U/10$ in \eqref{tnd}, i.e., imposing a dependency of the mode $\nu$ on the matrix density, allows us to recover the values of both the mean speed and the effective speed, without imposing any of them, as reported in Supplementary Table \ref{V_Gaussian}. 

\begin{figure}[!h]
\begin{center}
\includegraphics[width=0.3\textwidth]{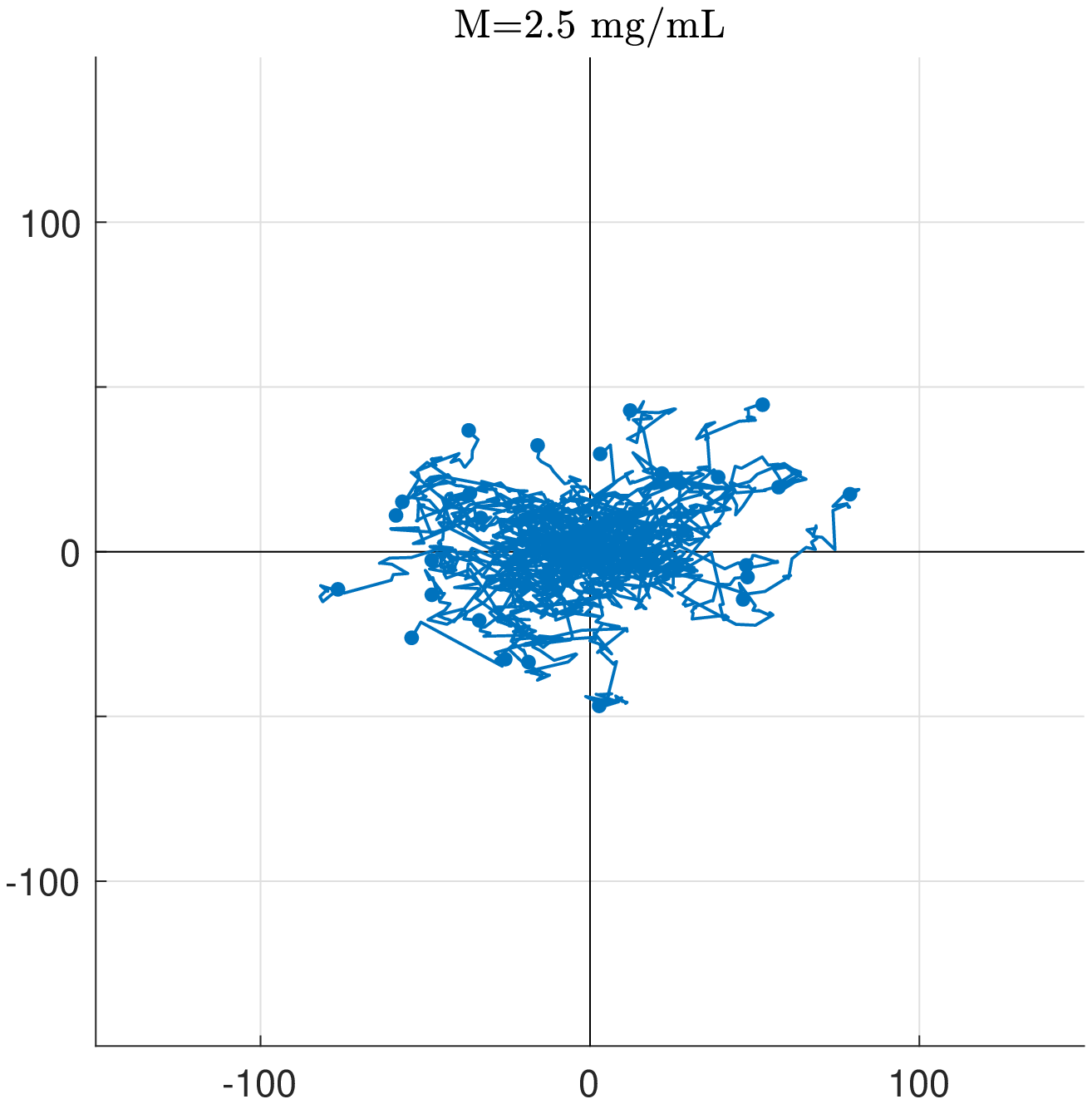}\hspace{0.3cm}
\includegraphics[width=0.3\textwidth]{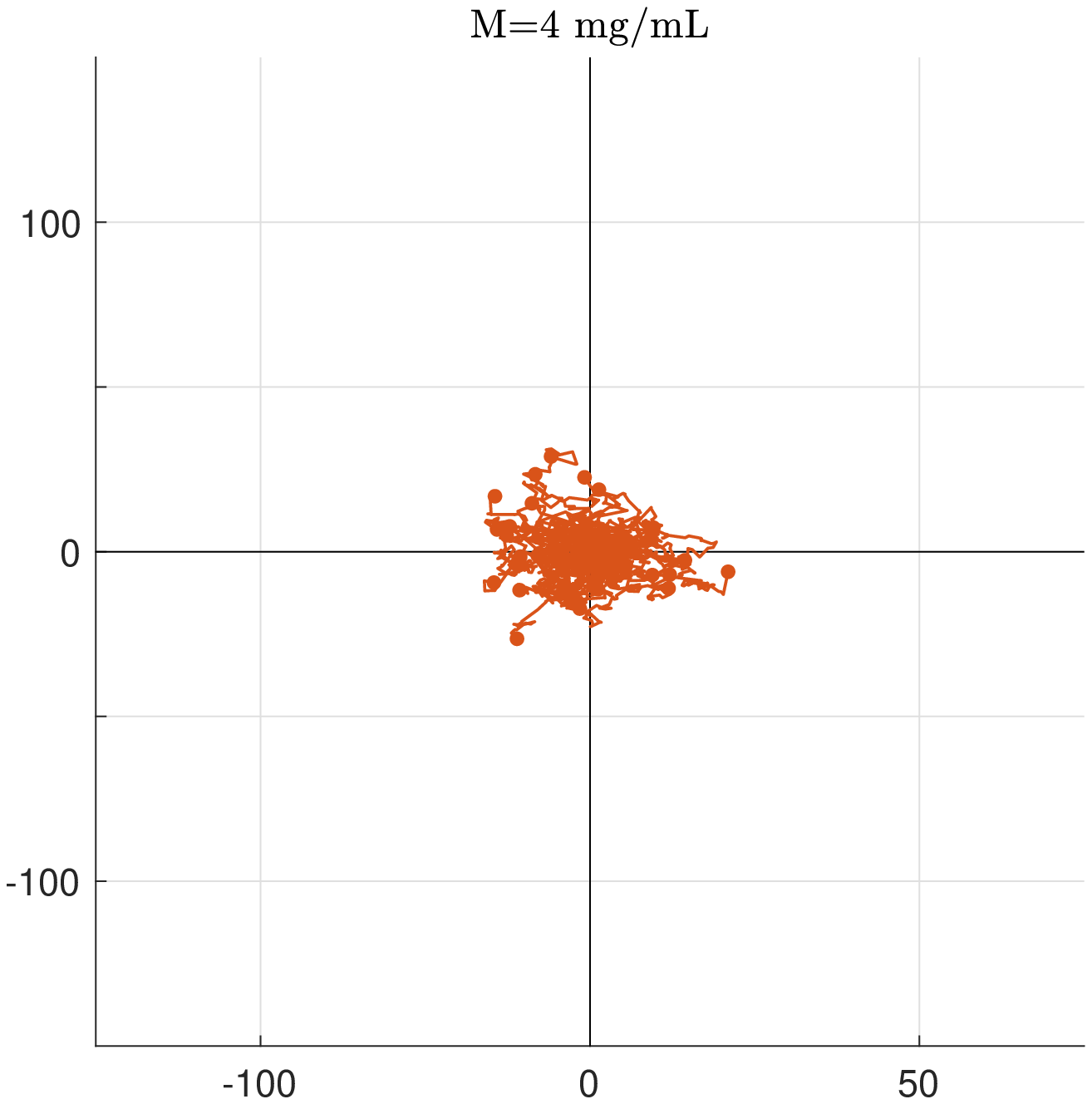}\hspace{0.3cm}
\includegraphics[width=0.3\textwidth]{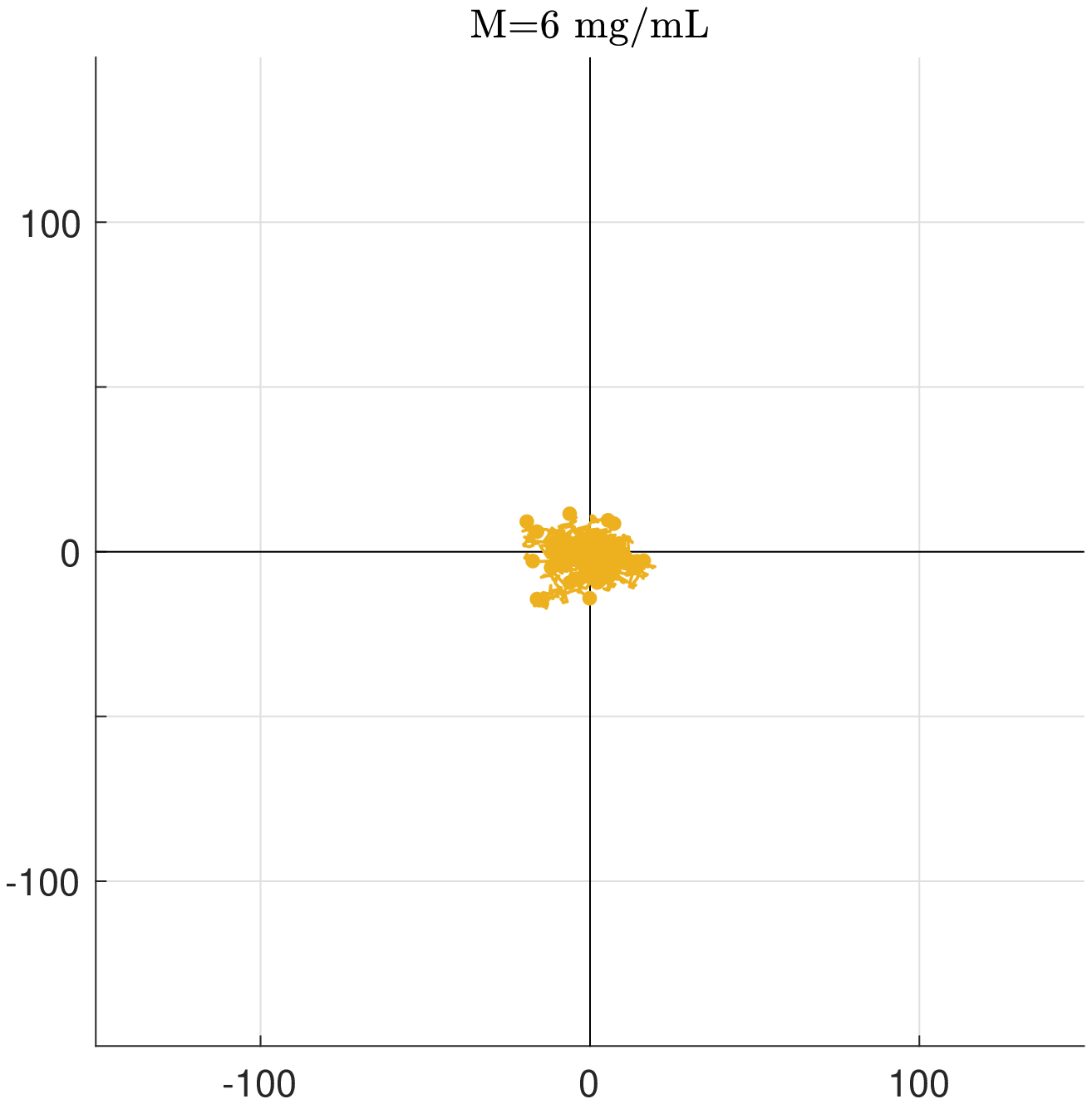}\hspace{0.3cm}
\caption{\textbf{Test 1.} Cell tracking in the domain $\Omega$ for the case $iii)$ of aligned fibers in the direction $\theta_q=0$ is shown, assuming a truncated Gaussian distribution for the speed. The parameter $k$ describing the alignment strength is here set to $k=1.2$. Simulations are run for $24$ h with $\Delta t=0.0167$ h.}
\label{fig_track_align}
\end{center}
\end{figure}

\noindent Looking at the cell tracking graphs reported in \cite{Plou}, we observe a clear difference in the cell spreading when the value of $M$ increases, showing a reduced motility for higher values of the matrix density, and a greater spreading along the horizontal direction. Only choosing a non-uniform speed distribution (cases $ii)$ and $iii)$) allows to recover the reduced motility for higher values of the ECM density (results related to case $ii)$ are shown in the Supplementary Fig. \ref{fig_track_noalign}). However, to reproduce cell alignment along a specific direction, a non-random description of the fiber network is necessary, i.e., case $iii)$. Fig. \ref{fig_track_align} shows the results of the cell tracking when fibers aligned along the direction indicated by $\theta_q=0$ are included. 

\noindent In conclusion, we have seen that the mere dependence of the frequency on the ECM density $M$ is not enough for recovering the behavior observed experimentally in terms of mean speed and MSD. We need to impose an $M$-dependent speed distribution to recover the appropriate MSD and effective speed. Moreover, if the probability density function $\psi$ is a truncated Gaussian distribution, we also recover the mean speed. We can observe a comparable directionality between the tracking when also an aligned fiber network is included. 

\subsection{The influence of steric hindrance on human breast cancer cell migration}\label{sec.num.2}
\begin{figure}[!h]
\centering
\includegraphics[width=0.45\textwidth]{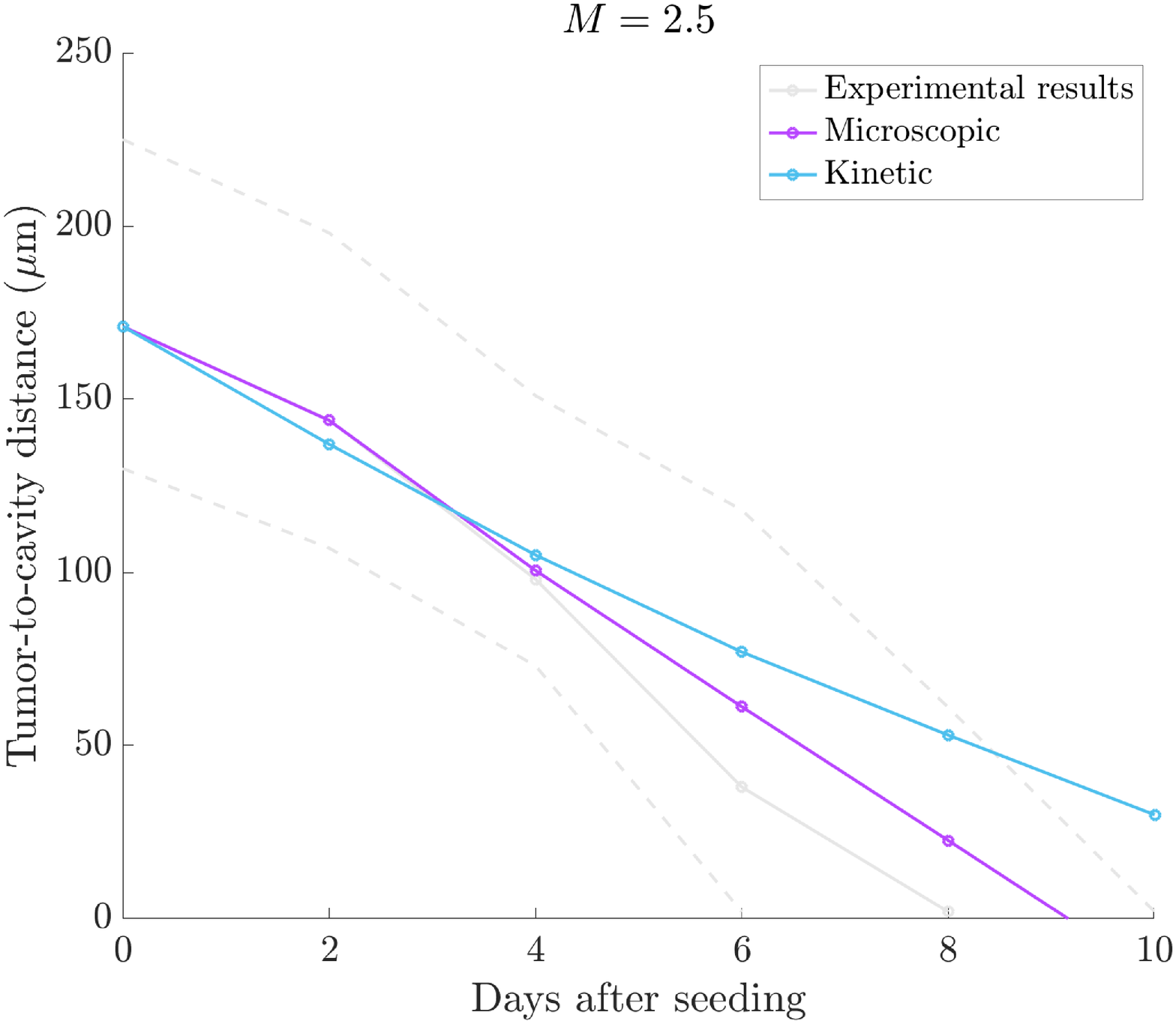}
\includegraphics[width=0.45\textwidth]{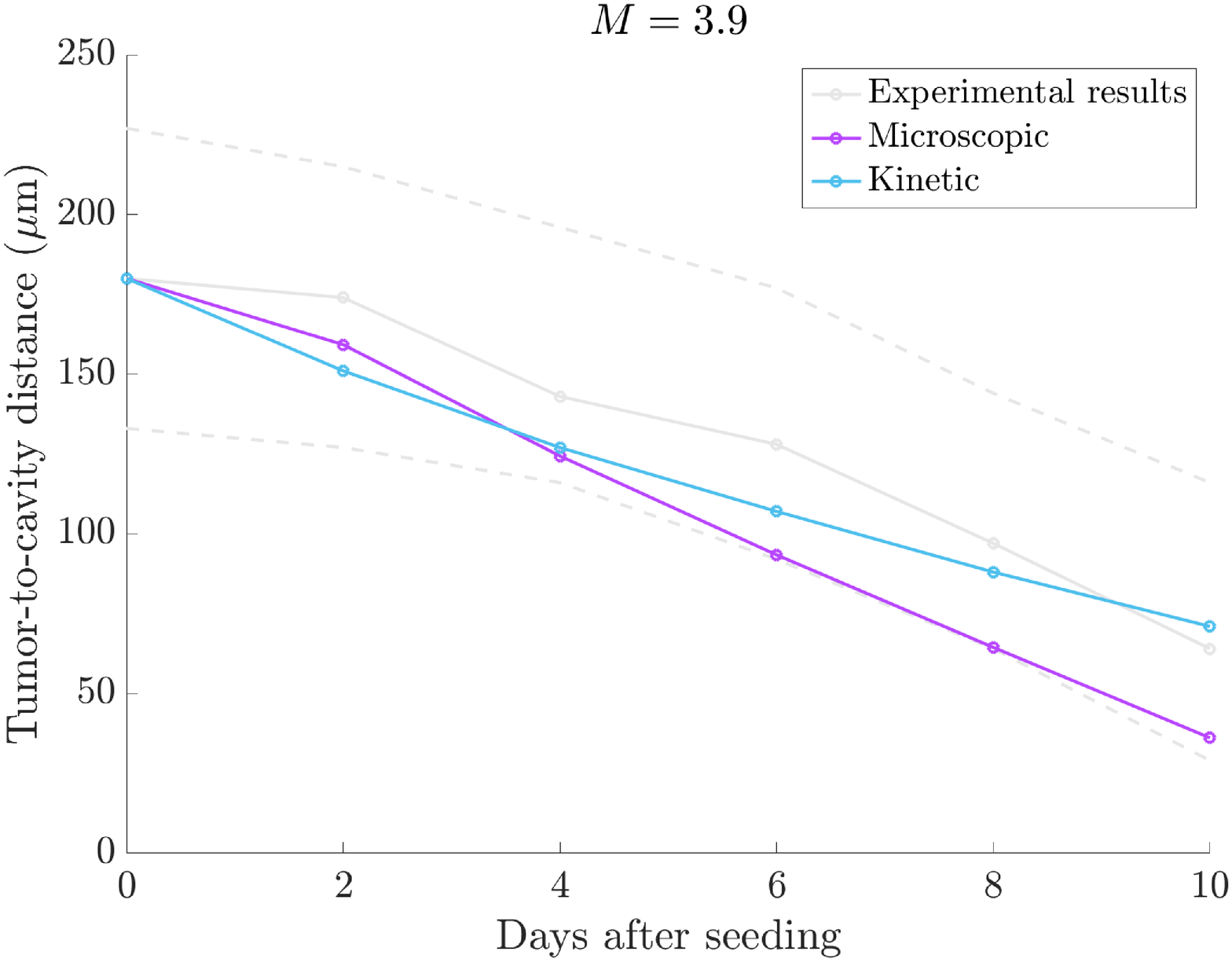}
\caption{\textbf{Test 2.} Tumor-to-cavity distances as a function of time, in low-density (left plot) and high-density (right plot) collagen. Light grey solid lines represent median values of the experimental setting, while light gray dashed lines represent 25th and 75th percentiles. Magenta solid lines represent median values of the microscopic numerical experiments, while blue solid lines represent the results of the kinetic model. We set $\mu=0.0003\,1/min$ and the mean speed $\bar{U}=0.0166\, \mu m/min$ for M$=2.5\, mg/mL$, while $\bar{U}=0.0137\, \mu m/min$ for M$=3.9\, mg/mL$. Simulations are run for 10 days with $\Delta t=7\cdot 10^{-3}$ days.}
\label{fig_BC_micro}
\end{figure}

We now consider the experimental results obtained in \cite{tien2020}, where the authors investigate how the physical properties of the ECM affect cancer cells' escape and invasion, using a microfluidic-based strategy (similar to the experiments in \cite{Plou}) on human breast cancer cells within a type I collagen gel. This culture model was developed to observe the invasion of breast cancer cells from an aggregate into a collagen gel under interstitial flow, which mimics the initial stage of breast tumor progression. Precisely, they consider tumors that are formed adjacent to empty cavities (mimicking the blind end of a lymphatic vessel) and observe how tumor cell behavior changes in response to different ECM density values. By altering the stiffness, the pore size - and hence the density of the collagen gel - and the magnitude of the interstitial flow through the gel, they find that the pore size is the main physical factor that determines the rate at which cells escape from their initial aggregate and invade the cavity. In particular, the movement of cancer cells through the collagen for two different collagen densities has been tracked over a period of 16 days, showing how lower collagen concentration promotes a faster tumor escape towards the empty cavity.\\
\noindent We focus on the results concerning tumor-to-cavity movement in low and high density collagen matrices, trying to reproduce the temporal evolution of the distance between the tumor and the cavity using the experimental setting proposed in \cite{tien2020}. We consider a domain ${\Omega=[0, 1000]\times[0, 600] \,\mu m^2}$. On the left side of the domain, between $x=0$ and $x=600 \,\mu m$, we locate 400 cells of radius $R=15 \, \mu m$, while on the right side we consider the empty cavity at a distance of $171\, \mu m $ or $180\, \mu m$, in the case M$=2.5\, mg/mL$ or M$=3.9\, mg/mL$, respectively, from the tumor cells. In the two cases the experimental mean speed reported in \cite{tien2020} are $0.0166$ and $0.0137$ $\mu m /min$. We consider the microscopic model \eqref{eq:micro_stoch}-\eqref{eq:space_ev}-\eqref{def:Povzner} with the choice \eqref{eq:micro_trans} (with $B=1$ as we are on a spatially homogeneous setting) where $\psi$ is given by \eqref{uvm} and $q$ is the unimodal von Mises distribution with $\theta_q=0$. This choice of $q$ mimics the presence of an oriented interstitial flow. We set $k_\psi=0.5$ in order to recover the experimental results reported in \cite{tien2020}. \\
\noindent As done in \cite{tien2020} in the \textit{in vitro} experiment, we perform $n$ numerical simulations of this microscopic model, with $n=73$ for $M=2.5\, mg/mL$ and $n=55$ for $M=3.9\, mg/mL$, and for each of them we calculate the distance between the cavity and the tumor cells at 2, 4, 6, 8, and 10 days and, finally, we consider the median of the obtained values. Results of the microscopic simulations are shown in Fig. \ref{fig_BC_micro}. In particular, the distance is defined as the distance between the cavity and the first, closer to the cavity, cell of the advancing cell aggregate. 

\begin{figure}[!h]
\centering
\includegraphics[width=0.45\textwidth]{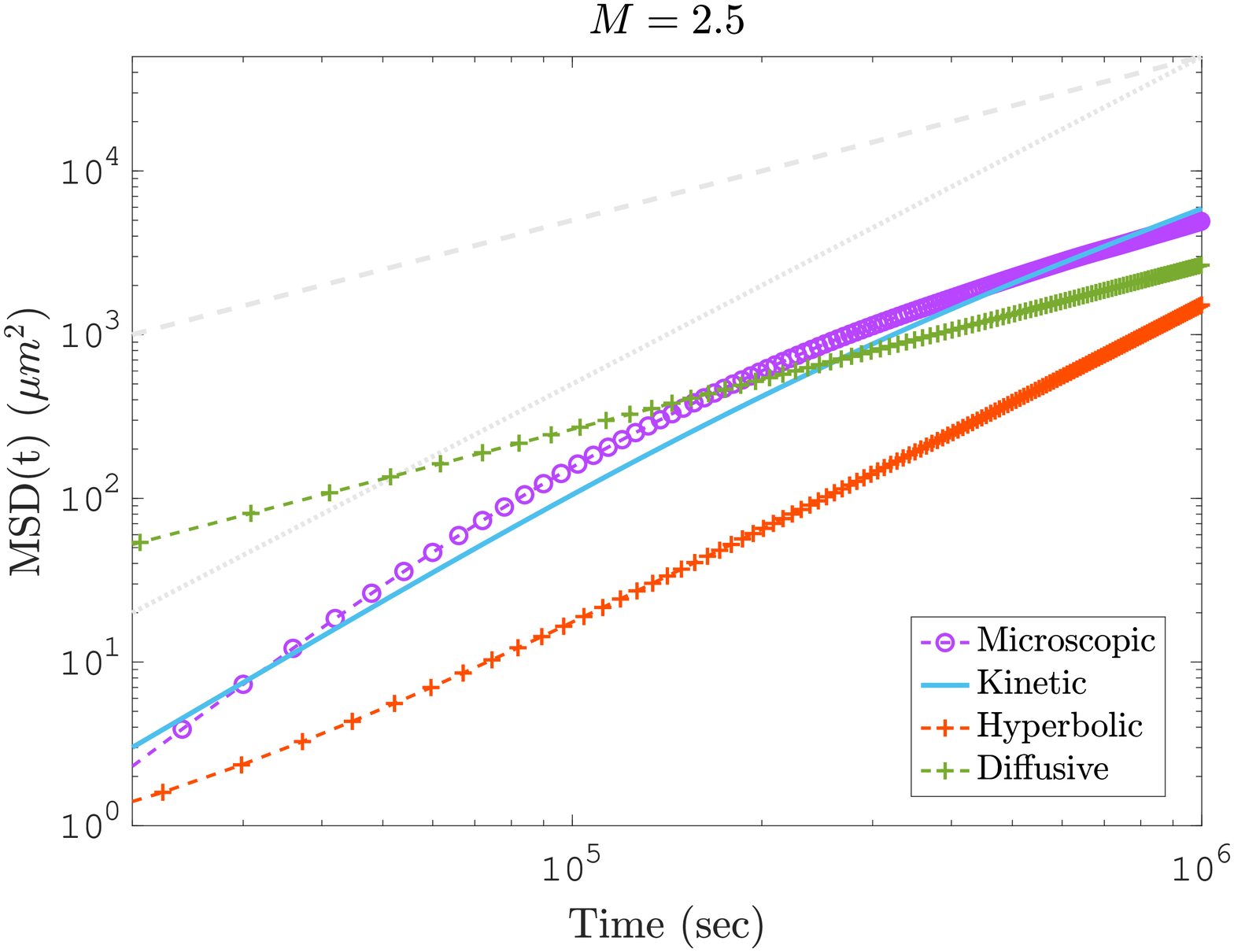}
\includegraphics[width=0.45\textwidth]{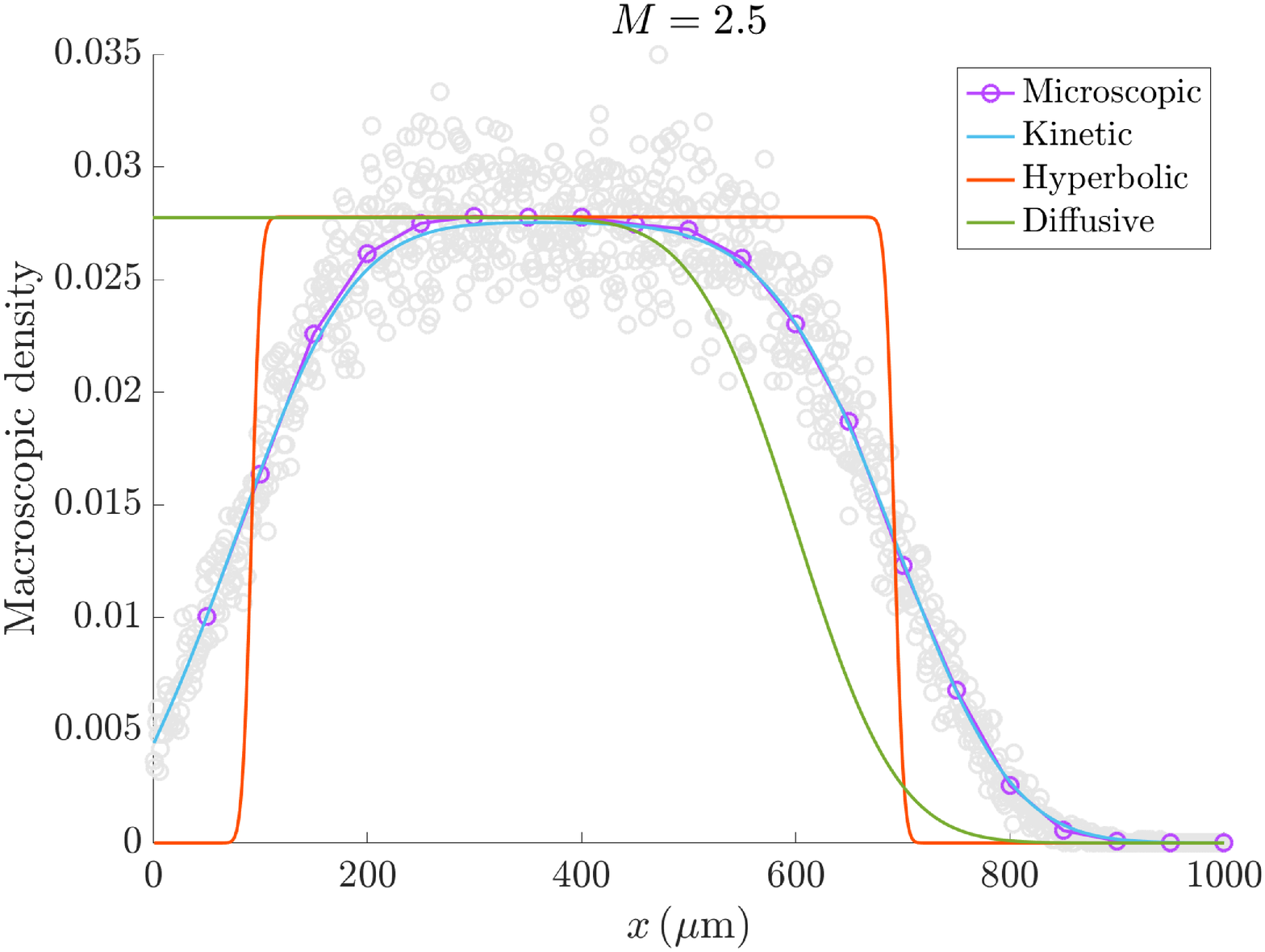}
\caption{\textbf{Test 2.} Mean squared displacement (MSD) and profile solutions for a fixed collagen density $M=2.5$. On the left, we compare the mean squared displacements prescribed by the microscopic (magenta), kinetic (blue), hyperbolic (orange), and diffusive (green) limit, respectively. We also report the lines $y=t$ (grey dashed), $y=t^2$ (grey dotted). On the right, we plot a one-dimensional section of the solution along the $x$-axis for the microscopic, kinetic, hyperbolic, and diffusive models, with the same choice of colors done for the MSD. In particular, for the microscopic model we build the corresponding histograms over both $20$ points in $[0, 1000]\mu$m (purple circles) and $1000$ points (light grey circles) and we construct the solution from them.}
\label{Profili}
\end{figure}

\noindent We observe how the microscopic model, with a unique alignment parameter $k$ is able to reproduce the trend in both collagen densities. The experimental results and the 25th and 75th percentiles are reported not for a direct comparison, but for showing that the difference in the rates of invasion in the two collagen densities is quite well reproduced.\\
\noindent In order to investigate more accurately the statistical evolution of the cells under the dynamics imposed by the microscopic model, we consider the kinetic model \eqref{eq:transp}-\eqref{eq:turning_operator}-\eqref{eq:trans1} and try to perform the same experiment. In this case, as we cannot track single cells, we need to impose a threshold for $\rho$ in order to compute the distance of the advancing cells aggregate from the cavity. In particular we choose $\rho_{th}=\dfrac{1}{400}=2.5 \cdot 10^{-3}$. The choice is arbitrary, but we observe that for a fixed threshold, the difference in the behavior of the cells in the two different collagen densities is in agreement with the experimental results. This, of course, corresponds to what we know from the theory, as the kinetic model \eqref{eq:transp}-\eqref{eq:turning_operator}-\eqref{eq:trans1} is derived from the microscopic model \eqref{eq:micro_stoch}-\eqref{eq:micro_trans}-\eqref{eq:space_ev}-\eqref{def:Povzner} in the limit $N \rightarrow \infty$ and $\Delta t \rightarrow 0^+$. This also shows the convenience of using the kinetic model instead of the microscopic one in order to obtain a complete statistical portrait with only one simulation, thus gaining a lower computational cost, instead of performing multiple simulations of the microscopic model or a simulation with a high $N$ and very small $dt$, which may be computationally challenging. In Fig. \ref{Profili} (right plot) we plot a section along the $x$ axis for a fixed $y\in [0, 600]\, \mu m$ of the macroscopic cell density $\rho(x,t)$, $x \in [0, 1000]\,\mu m$. 

\noindent In particular, we plot the solution of the kinetic model (in blue) that we recover from the definition \eqref{def_rho} and the solution of the microscopic model $\rho^{MC}(x,t)$ that we have run with $N=10^6$ particles and $\Delta t=0.001$ days. We construct the solution $\rho^{MC}(x,t)$ building the corresponding histograms both with $20$ points (purple circles) and 1000 points (light grey circles) over the interval $[0, 1000] \,\mu$m. We remark that there is an excellent agreement, as expected, between the solution $\rho^{MC}(x,t)$ of the microscopic model \eqref{eq:micro_stoch}-\eqref{eq:micro_trans}-\eqref{eq:space_ev}-\eqref{def:Povzner} and the solution $\rho(x,t)$ of the kinetic model \eqref{eq:transp}-\eqref{eq:turning_operator}-\eqref{eq:trans1}.  Always with the aim of reducing the computational effort, we look for the appropriate macroscopic model. To this aim, we observe the mean squared displacement shown in Fig. \ref{Profili} (left plot). For completeness, we look at both the microscopic and the kinetic model and we also report the lines $y=t$ (dashed grey) and $y=t^2$ (dotted grey) for direct comparison. We observe that the mean squared displacement prescribed by the microscopic (and, consequently, by the kinetic) model does not correspond to either a diffusive or a purely directed behavior. Therefore, a diffusive limit or a hyperbolic limit cannot be used for predicting accurately the behavior of the macroscopic quantities. As a consequence, in Fig. \ref{Profili} (right plot), the solution of the hyperbolic model \eqref{macro_hyp} (in orange) and the diffusive model \eqref{macro_diffdrift} (in green) done with $\varepsilon=10^{-3}$ do not reproduce accurately the solution of the kinetic model \eqref{eq:transp}-\eqref{eq:turning_operator}-\eqref{eq:trans1}.

\subsection{The interplay of steric hindrance and fiber alignment}\label{sec.num.3}
\begin{figure}[!h]
\centering
\includegraphics[width=0.35\textwidth]{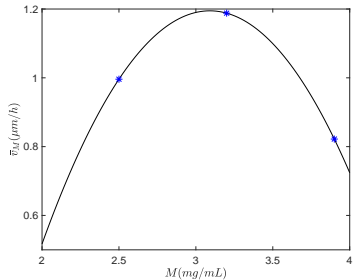}
\caption{\textbf{Test 3.} Mean cell speed depending on the density of the extracellular matrix. The blue star indicates the three different pairs of $(M,\bar{v}_M)$ used for the numerical experiments of this test.}
\label{barU_test3}
\end{figure}

We now use our model to investigate the interplay between steric hindrance and fiber alignment. Precisely, we analyze how cell migration on collagen fibers varies with respect to different combinations of matrix densities and fiber alignment. The main motivation for this test comes from a series of biological experiments that study separately the impact of stromal collagen concentration \cite{provenzano2008collagen} and collagen alignment \cite{provenzano2006collagen} on tumor local formation and invasion. In particular, in \cite{provenzano2006collagen} the authors use mouse breast tumor models to observe and define three tumor-associated collagen signatures (TACS), which are considered markers to locate and characterize tumor invasion. Specifically, TACS-1 refers to the presence of locally dense collagen randomly disposed fibers within the globally increased collagen concentration surrounding tumors, TACS-2 is defined as straightened collagen fibers stretched around the tumor and constraining its volume, while TACS-3 identifies radially aligned collagen fibers that facilitate local invasion. These observations allow the use of collagen alignment to quantify local invasion. Furthermore, in \cite{provenzano2008collagen}, the authors extend the analysis, looking at the influence of the extracellular matrix on breast carcinoma development using a tumor model with increased stromal collagen in mouse mammary tissue. They demonstrate how this increased collagen, coupled with the different collagen-associated signatures, significantly increases tumor formation and results in a more invasive phenotype. Directed cell migration by contact guidance in aligned collagenous ECM has been also observed in \cite{Ray2017BioJ}, where the authors propose a method to align collagen gels that provides a controlled microenvironment for in vitro experiments. They quantify breast cancer cell behavior in these anisotropic constructs, showing how motility is enhanced in aligned collagen matrices and for a subpopulation of carcinoma cells, namely cancer stem cells (CSCs). In particular, these cells are characterized by smaller cell size and a high degree of phenotypic plasticity which makes them more able to adapt to contact-guided migration.\\
We focus on the results in \cite{Ray2017BioJ} concerning cell motility with respect to the alignment of the fibers and the density of the ECM. In particular, for our analysis we translate the differences in the cell size between cancer cells and CSCs as a difference in the matrix pore size, meaning that we expect to observe enhanced migration in less dense regions (where the pore size is bigger). We consider the domain ${\Omega=[-150, 150]\times[-150, 150] \mu m^2}$ with an initial Gaussian distribution of cells centered in ${(x_0,y_0)=(0,0)}$ and of variance $\sigma_0^2=10^{-4}$. We analyze different scenarios which combine three possible values for the matrix density, meaning $M=2.5$, $3.2$, or $3.9 \,mg/mL$, and three possible values of the parameter $k$ responsible for the strength of fiber alignment, i.e., $k=0$, $5$, or $10$. In particular, we consider a spatially homogeneous setting where $q$ is the bimodal von Mises distribution \eqref{bvm} with $\theta_q=0$, while for the speed distribution we consider $\psi$ given in \eqref{uvm}, where we set the value of the maximum cell velocity $U=3.34 \mu m/h$ and the concentration parameter $k_\psi=10$, while the cell mean speed is given as a function of the ECM density $\bar{v}_M=\bar{v}_M(M)$ as shown in Fig. \ref{barU_test3}. 

\begin{figure}[!h]
\centering
\includegraphics[width=0.45\textwidth]{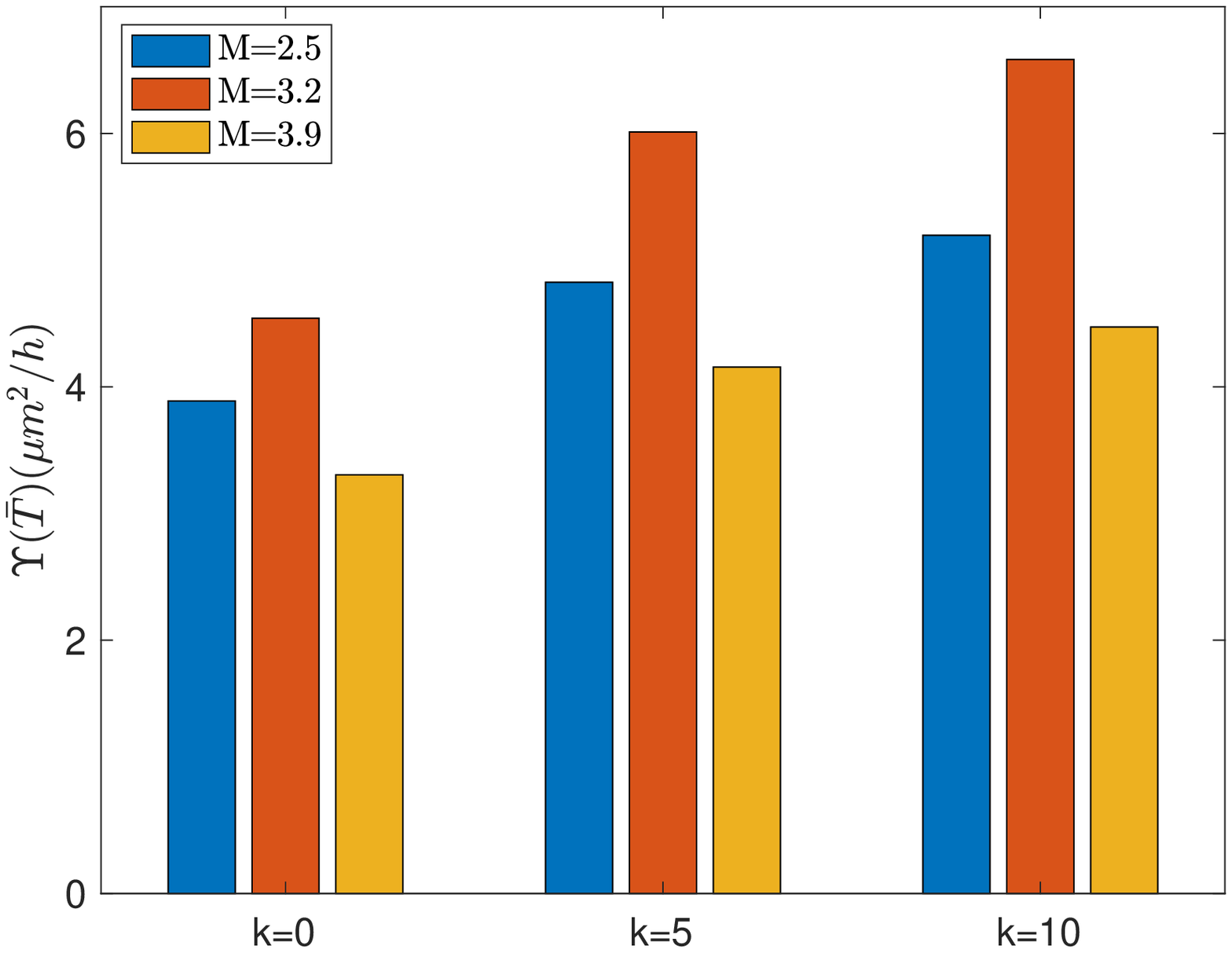}
\includegraphics[width=0.45\textwidth]{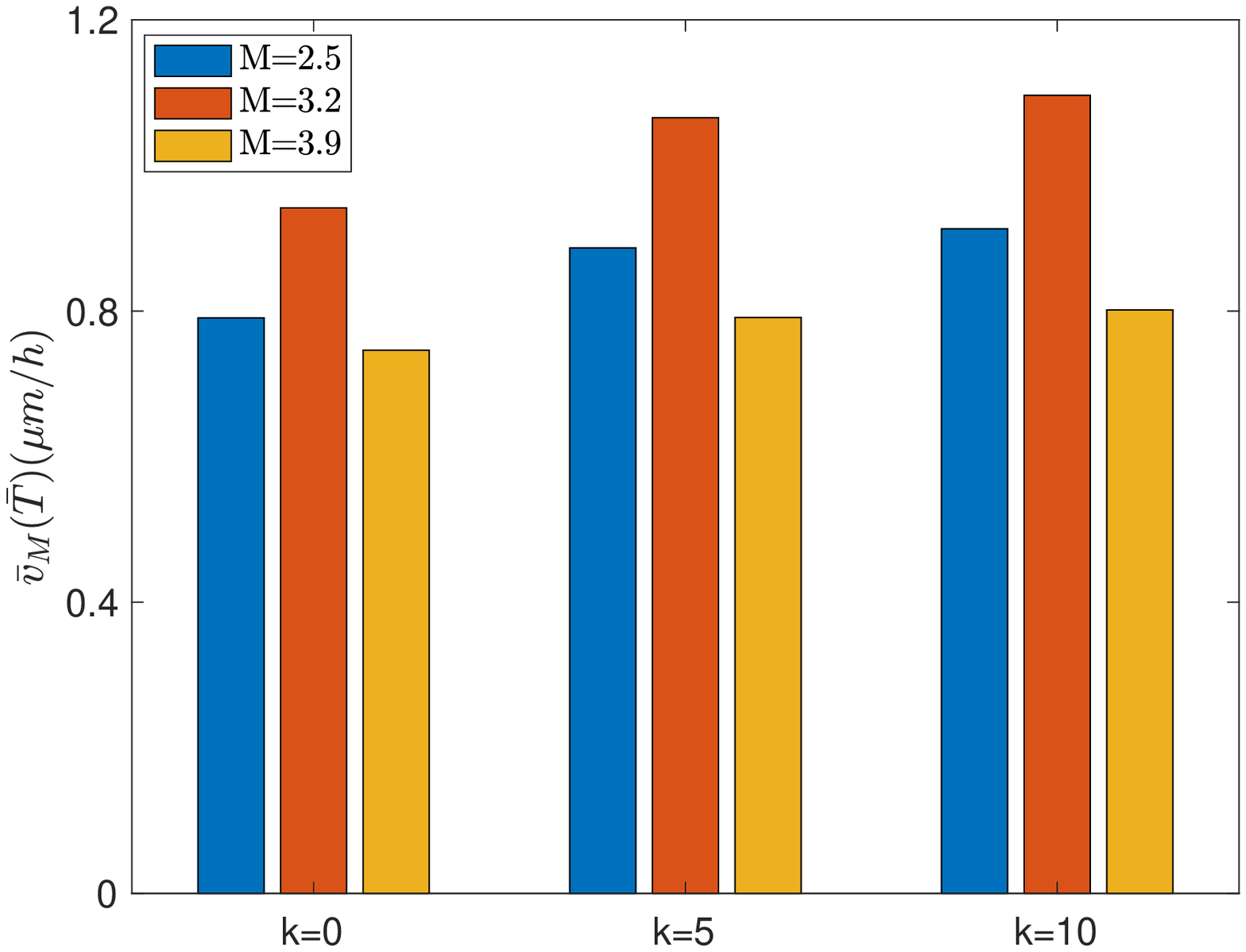}
\caption{\textbf{Test 3.} Cell motility ($\Upsilon$) and cell mean speed ($\bar{v}_M$) in the direction $\theta_q=0$ at $\bar{T}=16 \,h$ for three different values of the collagen density $M=2.5, 3.2, 3.9 \, mg/mL$ and three different values of the fiber alignment strength $k=0,5,10$.}
\label{MatrixAllign}
\end{figure}

\noindent In particular, we consider three different values of the ECM density as given in \cite{tien2020} and the corresponding mean speeds. The values of the mean speeds are marked by the three blu stars in Fig. \ref{barU_test3} and the black straight line represents the interpolating polynomial of degree two that may be considered to approximate the behavior of the mean speed as a function of the ECM density \cite{Wolf_Friedl.13}. We remark that, as in \cite{tien2020}, we have an optimal matrix density corresponding to the maximum possible speed, while for smaller values of the ECM density the speed is lower, because this corresponds to larger pores and to less efficient cell migration, as shown in \cite{Wolf_Friedl.13}. The value of the mean speed also decreases for higher values of the ECM density because of the physical limit of migration effect, as reported also in \cite{Wolf_Friedl.13}. We study the effects of matrix density and alignment on cell mean speed $\bar{v}_M$ and cell motility $\Upsilon$ in the direction of the alignment, defining cell motility as 
\[
\Upsilon(t):=\dfrac{MSD(t)}{t}\,.
\]
Following \cite{Ray2017BioJ}, we compare the values of cell motility and mean speed after $\bar{T}=16\, h$. Results of the simulations of the kinetic model \eqref{eq:transp}-\eqref{eq:turning_operator}-\eqref{eq:trans1} in this setting are shown in Fig. \ref{MatrixAllign}.

\noindent In agreement with the results in \cite{Ray2017BioJ}, we observed how, for the same value of collagen density $M$, a stronger alignment of the fibers enhances the cell mean speed along the fiber tracks and this determines an increased motility in the same direction, compared with the control case $k=0$. In particular, the differences in both $\bar{v}_M$ and $\Upsilon$ are more evident for $M=2.5 \, mg/mL$ and $3.2\, mg/mL$, while for $M=3.9 mg/mL$ the matrix density seems to be a greater obstacle for cell migration, even in strongly aligned environment. Comparing, instead, the cell behavior for the fixed value of the fiber alignment, we notice how the results about both mean speed and motility are in accordance with the relation $\bar{v}_M=\bar{v}_M(M)$ illustrated in Fig. \ref{barU_test3} and we obtain a greater mean speed and motility for $M=3.2\, mg/mL$. In particular, comparing the left and right plots of Fig. \ref{MatrixAllign}, we notice that the effect of the alignment on the mean speed seems to be lower than its effect on the cell motility, as it was also observed in \cite{Ray2017BioJ}. In the interplay between contact guidance and steric hindrance, this shows a prominent role of contact guidance in the overall dynamics.

\section{Conclusion}
In this work, we have presented a mathematical model for the description of contact guidance and steric hindrance, two fundamental mechanisms related to cell migration on the ECM. This model relies on a non-local (in the physical space) sensing of the ECM that allows to take into account the role of cell protrusions, which may be extended up to several cell diameters, in heterogeneous environments. Considering the success of using kinetic models for describing cell motion, especially because of their genuine multiscale nature, we have transferred existing mathematical methodologies of kinetic equations for multi-agent systems, already widely applied in other fields \cite{Chiarello2022unp,pareschi2013BOOK}, to the context of cell migration. Starting from microscopic discrete in time stochastic processes, which also involve non-local aspects, we have accurately described the microscopic dynamics and, then, rigorously derived a kinetic model implementing the chosen dynamics in the form of a collision-like Boltzmann equation. In particular, we have established the parallelism between this class of models and the velocity-jump processes, already commonly used to describe cell migration. This has allowed us not only to give a more detailed microscopic description of the considered dynamics, but also to obtain a microscopic algorithm for simulating them and, thus, performing in silico experiments. 

\noindent The kinetic model that we have formally derived gives the complete statistical description of the studied system and it implements exactly the prescribed microscopic dynamics, instead of postulating them at the mesoscopic scale. Moreover, this kinetic formulation allows to run a unique simulation instead of performing multiple independent simulations. 

\noindent We have shown how to derive from the mesoscopic level the macroscopic models in different regimes according to the parameters of the system, which are leading the different involved phenomena. In particular, this has allowed us to show how not always diffusive or advective models can be reliable in order to make accurate predictions and, thus, models directly stated at the macroscopic level could not be able to correctly describe cell behaviors. 

\noindent We have applied our model to the study of steric hindrance and contact guidance in several scenarios related to breast cancer dissemination, which is a prominent medical issue. In particular, many efforts have been done, especially in the medical and biological community, for the study of this process, but most of the difficulties are still encountered in the design of effective experimental platforms. Thus, our approach aims at providing a useful platform for performing in silico experiments. To this aim and especially for what concerns steric hindrance, we have firstly validated our microscopic model studying the minimal and essential mechanisms that should be included at the microscopic level in order to retrieve several experimental results \cite{Plou,tien2020}. Then, we have performed further experiments that couple the two mechanisms, meaning steric hindrance and contact guidance. This has allowed us to make some predictions on the behavior of cells that undergo both mechanisms. In particular, the obtained results can be actually supported by biological evidence.

\noindent For completeness, we have to highlight that we have focussed our attention on breast cancer dissemination, but the two considered mechanisms are involved in many other processes. Thus, except for appropriate modifications, our methodology could be adapted to the description of other phenomena. In particular, the microscopic dynamics could be modified and enriched, for instance modelling an independent sensing of multiple directional cues, in order to describe other observed microscopic mechanisms and to derive a proper mesoscopic and macroscopic description. We also remark that the established parallelism with collisional Boltzmann equations can be used in order to inherit all the analytical tools that are widely used in the community of multi-agent systems and kinetic equations. In particular, it is also used in order to tackle technical difficulties that may be encountered in more complex models implementing velocity-jump processes, such as the determination of stationary equilibria and the derivation of macroscopic limits.

\section*{Acknowledgments}
The authors would like to thank Prof. Luigi Preziosi for fruitful discussions and valuable comments. 

\bibliographystyle{siamplain}
\bibliography{references}

\end{document}

\title{Supplementary Material: A non-local kinetic model for cell migration: a study of the interplay between contact guidance and steric hindrance}
\author{Martina Conte, Nadia Loy}
\date{\today}
\maketitle
\section{Macroscopic Limit }\label{Macro_model_SM}
To analyze the macroscopic behavior of the system and, thus, its overall trend, we derive the macroscopic evolution equation for $\rho(t,\x)$ in the emerging regime. As explained in Section \ref{Macro_model}, the appropriate scaling results from a proper non-dimensionalization of the system based on the introduced small parameter $\epsilon \ll 1$. The choices of performing a parabolic scaling ($\tau=\epsilon^2t$) or a hyperbolic scaling ($\tau=\epsilon t$) can be done by  measuring the mean squared displacement (MSD), where $\tau$ refers to the macroscopic time scale introduced in \eqref{eq:diff_scale}. 

\noindent Diffusive and hydrodynamic limits  techniques relying on Hilbert expansions for transport equations with velocity jump processes have been widely treated in \cite{Hillen.05,Othmer_Hillen.00, loy2019JMB, Loy_Preziosi2,Othmer_Hillen.02}. 

\noindent We consider a framework in which, up to the spatial scaling \eqref{eq:scale_space}, we can expand the transition probability as
\[
T(\boldsymbol{\xi},v,\hv)=T_0(\boldsymbol{\xi},v,\hv)+\epsilon T_1(\boldsymbol{\xi},v,\hv)+\mathcal{O}(\epsilon^2).
\]
This means that there are different orders of bias. If we assume that $\mu=\mathcal{O}(1)$, we denote by $\mathcal{J}^0$ and $\mathcal{J}^1$ the corresponding operators defined by $T_0$ and $T_1$ respectively, and
we assume that
\begin{equation}\label{diff.cond.1}
\displaystyle \int_0^U \int_{\mathbb{S}^{d-1}} T_0(\boldsymbol{\xi}, v, \hv)  \, d\hv\,dv  =1, \quad \int_0^U \int_{\mathbb{S}^{d-1}} T_i(\boldsymbol{\xi}, v, \hv)  \, d\hv dv   =0 \quad \forall i \geq 1 .
\end{equation}
The corresponding average and variance-covariance matrix are given by
\begin{equation}\label{UT_exp}
\Ub_T^i(\boldsymbol{\xi})=\int_0^U \int_{\mathbb{S}^{d-1}} T_i (\boldsymbol{\xi},v,\hv)\bv\, d\hv dv 
\end{equation}
and 
\begin{equation}\label{DT_exp}
\mathbb{D}_T^i(\boldsymbol{\xi})=\int_0^U \int_{\mathbb{S}^{d-1}} T_i(\boldsymbol{\xi}, v,\hv) (\bv-\Ub^i_T)\otimes (\bv-\Ub^i_T) \, d\hv dv \,.
\end{equation}
Considering the Hilbert expansion of the distribution function $p$ 
\begin{equation}\label{Hilbert_exp}
p=p_0+\epsilon p_1 +\mathcal{O}(\epsilon^2)\,,
\end{equation}
if there is conservation of mass, we have that all the mass is in $p_0$ \cite{Othmer_Hillen.00}, i.e.,
\begin{equation}\label{rho0}
\rho_0=\rho, \quad \rho_i=0 \quad \forall   i \geq 1 \, ,
\end{equation}
where ${\rho_i=\int_0^U \int_{\mathbb{S}^{d-1}} p_i \, d\hat{\bv}\,dv }$. 
Furthermore, for performing the diffusive limit we shall assume that
 ${\int_0^U \int_{\mathbb{S}^{d-1}} p_i \,\bv\, d\hat{\bv}\,dv  =0 } \ \forall i \geq 2$ \cite{Othmer_Hillen.00}.
The fundamental property for performing the diffusive limit requires
\begin{equation}\label{UT0.0_SM}
\Ub^0_T=0,
\end{equation}
meaning that the leading order of the drift vanishes. This is coherent with the fact that the time scale $\tau=\epsilon^2 t$ is chosen because macroscopically the phenomenon is diffusion-driven.
In case of $T$ given by \eqref{eq:trans1}, this typically happens when the sensing radius is small \cite{loy_conte2020,loy2019JMB}.

\noindent Equation \eqref{eq:transp}-\eqref{eq:turning_operator}, rescaled according to \eqref{eq:scale_space}-\eqref{eq:diff_scale} with $\gamma=1$, reads
\begin{equation}
\begin{aligned}
\epsilon^2\dfrac{\partial p}{\partial \tau}(\tau,{\bf \boldsymbol{\xi}},v,\hv) + \epsilon\bv\cdot \nabla p(\tau,{\bf \boldsymbol{\xi}},v,\hv)=&\mathcal{J}^0[p]+\epsilon\mathcal{J}^1[p] +\mathcal{O}(\epsilon^2) \, .
\end{aligned} 
\end{equation}
Equating the terms of equal order in $\epsilon$, we obtain the following system of equations.\\
In $\epsilon^{0}$:
\begin{equation}
\mathcal{J}^0[p_{0}](\boldsymbol{\xi},v,\hv)\equiv  \eta  \Big( \rho_0 T_0({\bf \boldsymbol{\xi}},v,\hv)-p_{0}(\tau,{\bf \boldsymbol{\xi}},v,\hv)  \Big)=0
\label{eps0q}
\end{equation}
In $\epsilon^{1}$:
\begin{equation}
\begin{aligned}
\nabla\cdot\big( p_{0}(\tau, {\bf \boldsymbol{\xi}},v,\hv)\bv\big)&=\mathcal{J}^0[p_1](\tau,{\bf \boldsymbol{\xi}},v,\hv) +\mathcal{J}^1[p_0](\tau,{\bf \boldsymbol{\xi}},v,\hv)=\\[8pt]
&=\eta  \big(\rho_1 T_0({\bf \boldsymbol{\xi}},v,\hv) -p_1(\tau,{\bf \boldsymbol{\xi}},v,\hv)\big)
+ \eta  \rho_0 T_1({\bf \boldsymbol{\xi}},v,\hv) 
\end{aligned}
\label{eps1q}
\end{equation}
In $\epsilon^{2}$:
\begin{equation}
\begin{array}{lclr}
&\dfrac{\partial }{\partial \tau}p_{0}(\tau,{\bf \boldsymbol{\xi}},v,\hv)+\nabla\cdot\big(p_{1}(\tau,{\bf \boldsymbol{\xi}},v,\hv)\bv \big)=\\[0.5cm]
&=\mathcal{J}^0[p_2]({\bf \boldsymbol{\xi}},v,\hv) +\mathcal{J}^1[p_1](\tau,{\bf \boldsymbol{\xi}},v,\hv)+\mathcal{J}^2[p_0](\tau,{\bf \boldsymbol{\xi}},v,\hv).
\label{eps2q}
\end{array}
\end{equation}
Eq. ($\ref{eps0q}$) implies 
\begin{equation}
p_0(\tau, {\bf \boldsymbol{\xi}},v,\hv)=\rho_0(\tau,{\bf \boldsymbol{\xi}})T_0({\bf \boldsymbol{\xi}},v,\hv)
\end{equation}
that is the equilibrium state of order zero.
From Eq. ($\ref{eps1q}$),  by inverting $\mathcal{J}^0$, we get
\begin{equation}
p_{1}({\bf \boldsymbol{\xi}},v,\hv)=-\dfrac{1}{\eta}  \nabla \cdot \big(\bv p_{0}\big)+  \rho_0 T_1({\bf \boldsymbol{\xi}},v,\hv) \, .
\end{equation}
Precisely, the functional solvability condition necessary for inverting $\mathcal{J}^0$ is
\begin{equation}\label{solv.1}
\int_0^U \int_{\mathbb{S}^{d-1}}\left[- \nabla \cdot \dfrac{1}{\eta} \big(\bv p_{0} \big)+ \rho_0 T_1({\bf \boldsymbol{\xi}},v,\hv)\right] \,d\hv\,dv =0 \!\!\!\!\qquad \textit{for}\,\,\,a.e.\,\,\, \boldsymbol{\xi},
\end{equation}
 which is satisfied because \eqref{UT0.0_SM} and the second of \eqref{diff.cond.1} are satisfied.
Integrating \eqref{eps2q} over $[0,U]\times \mathbb{S}^{d-1}$, we obtain the macroscopic diffusive limit given by (dropping the dependencies)
\begin{equation}\label{macro_diff_SM}
\dfrac{\partial}{\partial {\tau}} \rho +\nabla \cdot \left( \Ub_T^1 \rho\right)=\nabla \cdot \left[ \dfrac{1}{\eta} \nabla \cdot \left(\mathbb{D}_T^0 \rho\right) \right]\,,
\end{equation}
being $\mathbb{D}_T^0$ 
the diffusion motility tensor and recalling \eqref{rho0}.
Equation \eqref{macro_diff_SM} is a diffusion-advection equation, where $\Ub_T^1$ is the drift velocity of first order.

\noindent If \eqref{UT0.0_SM} does not hold, as typically happens if $R$ is large with respect to the length of variation of the external field $m$,
but the non-dimensionalization of the system or experimental observations prescribe a diffusive regime,  we can consider a drift-diffusion limit as it was done in \cite{loy2021EJAM}. Setting $p(\tau,\bxi,\bv)=u(\tau,z,\bv)$, with $z=\bxi-\bU_T\tau$, we have
\[
 \dfrac{\partial}{\partial {\tau}} p+\bv\cdot\nabla p=\mathcal{L}[p]\,\,\Longleftrightarrow\,\,\dfrac{\partial}{\partial {\tau}}  u+\nabla\cdot((\bv-\bU_T)u)=\mathcal{L}[u]
\]
Here, a parabolic scaling can be applied, getting
\[
\varepsilon^2\dfrac{\partial}{\partial {\tau}}  u+\varepsilon\nabla\cdot((\bv-\bU_T)u)=\mathcal{L}[u]\,.
\]
Expading $u=u_0+\varepsilon u_1+O(\varepsilon^2)$ and plugging it into the scaled equation, we can collect the different power of $\varepsilon$:
\begin{equation}
\varepsilon^0:\,\,\,\mathcal{L}[u_0]=0\Longrightarrow u_0=\bar{u}_0(\tau,z)T
\end{equation}
with $\bar{u}_0=\int_0^U \int_{\mathbb{S}^{d-1}}u_0 \, d\hv\,dv$.
\begin{equation}
\varepsilon^1:\,\,\,\mathcal{L}[u_1]=\nabla\cdot((\bv-\bU_T)\bar{u}_0T)
\end{equation}
whose solvability condition is clearly satisfied since
\[
\nabla\cdot \int_0^U \int_{\mathbb{S}^{d-1}}((\bv-\bU_T)\bar{u}_0T) \, d\hv\,dv=0\,.
\]
Inverting $\mathcal{L}$ we get
\[
u_1=-\dfrac{1}{\eta}\nabla\cdot[(\bv-\bU_T)\bar{u}_0T]\,.
\]
Finally,
 \begin{equation}
\varepsilon^2:\,\,\,\dfrac{\partial}{\partial {\tau}} u_0+\nabla\cdot((\bv-\bU_T)u_1)=\mathcal{L}[u_2]\,.
\label{eq_eps2}
\end{equation}
Now, the solvability condition reads
\begin{equation}
\nabla\cdot\int_0^U \int_{\mathbb{S}^{d-1}}(\bv-\bU_T)u_1d\bv
=-\nabla\cdot\left[\dfrac{1}{\eta}\int_0^U \int_{\mathbb{S}^{d-1}}(\bv-\bU_T)\nabla\cdot[(\bv-\bU_T)\bar{u}_0T] \, d\hv\, dv \right]\,.
\end{equation}
Recalling the expression of $\mathbb{D}_T$, we can write
\begin{equation*}
 \begin{split}
\nabla\cdot(\mathbb{D}_T\bar{u}_0)&=\nabla\cdot\left(\int_0^U \int_{\mathbb{S}^{d-1}}(\bv-\bU_T)\otimes(\bv-\bU_T)T\bar{u}_0 \, d\hv \, dv \right)\\[0.2cm]
&=\int_0^U \int_{\mathbb{S}^{d-1}}(\bv-\bU_T)\nabla\cdot[(\bv-\bU_T)T\bar{u}_0] \,d\hv\, dv \\
&+\int_0^U \int_{\mathbb{S}^{d-1}}[(\bv-\bU_T)T\bar{u}_0]\cdot \nabla(\bv-\bU_T) \, d\hv\, dv \\[0.2cm]
&=\int_0^U \int_{\mathbb{S}^{d-1}}(\bv-\bU_T)\nabla\cdot[(\bv-\bU_T)T\bar{u}_0] \,d\hv\, dv .
 \end{split}
\end{equation*}
Thus, the solvability condition on \eqref{eq_eps2} reads
\[
\dfrac{\partial}{\partial {\tau}} \bar{u}_0-\nabla\cdot\left(\dfrac{1}{\eta}\nabla\cdot(\mathbb{D}_T\bar{u}_0)\right)=0\,.
\]
Going back to the original variable $p$ and remembering \eqref{rho0}, we get
\begin{equation}\label{macro_diffdrift_SM}
\dfrac{\partial}{\partial {\tau}} \rho+\nabla\cdot(\bU_T\rho)=\nabla\cdot\left(\dfrac{1}{\eta}\nabla\cdot(\mathbb{D}_T\rho)\right)\,.
\end{equation}

\noindent If, instead, a hyperbolic scaling is required, we can use the results presented in \cite{Hillen.05} that gives
\begin{equation}\label{macro_hyp_SM}
 \dfrac{\partial}{\partial {\tau}} \rho + \nabla\cdot (\rho \bU_T)=\varepsilon\nabla\cdot\left(\dfrac{1}{\eta}\nabla\cdot(\mathbb{D}_T\rho)+\dfrac{1}{\eta}\rho\,\bU_T\nabla\cdot\bU_T\right)\,.
\end{equation}
This is the equation with the first-order correction in which we can recover the dependency on the ECM through the frequency $\eta$ in the correction term.

\section{Test 1: supplementary material}
Here we reported the Supplementary Tables built for {\bf Test 1} that show how using the unimodal von Mises distribution or the truncated Gaussian distribution it is possible to recover the experimental values of mean and effective cell speed. In particular, we indicate with $\bar{v}_{\text{exp}}$ and $\tilde{v}_{\text{exp}}$ the experimental mean and effective speed, respectively, while with $\bar{v}_{\text{num}}$ and $\tilde{v}_{\text{num}}$ the numerical ones.

\begin{table} [!ht]
\begin{center}
   \begin{tabular}{c|c|c} \hline
   \toprule  
   \rule{0pt}{0.5ex}M [$mg/mL$]&$\tilde{v}_{\text{exp}}$ & $\tilde{v}_{\text{num}}$\\
  \midrule
   \rule{0pt}{2ex}$2.5$ &0.0285& 0.0261\\
   \rule{0pt}{2ex}$4$ &0.0105& 0.0129 \\
   \rule{0pt}{2ex}$6$ &0.0079& 0.0068 \\
   \bottomrule
    \end{tabular}
\end{center}
\caption{\textbf{Test 1.} Effective cell speed for different collagen concentrations for both experimental and numerical settings obtained using the unimodal von Mises distribution \eqref{uvm} for the speed, imposing the mean speed values. The values the effective speed are expressed in [$\mu m/min$].}
\label{Veff_VM}
\end{table}

\begin{table} [!ht]
\begin{center}
   \begin{tabular}{c|c|c|c|c} \hline
   \toprule  
   \rule{0pt}{0.5ex}M [$mg/mL$] &$\bar{v}_{\text{exp}}$ & $\bar{v}_{\text{num}}$ & $\tilde{v}_{\text{exp}}$& $\tilde{v}_{\text{num}}$\\
  \midrule
   \rule{0pt}{2ex}$2.5$&0.1696 &0.1576 &0.0285& 0.0254\\
   \rule{0pt}{2ex}$4$ &0.104 &0.0997 &0.0105& 0.0133 \\
   \rule{0pt}{2ex}$6$ & 0.063 &0.0704 &0.0079& 0.0079 \\
   \bottomrule
    \end{tabular}
\end{center}
\caption{\textbf{Test 1.} Mean and effective cell speed for different collagen concentrations for both experimental and numerical settings obtained using the truncated Gaussian distribution \eqref{tnd} for the speed, imposing the value of $\sigma$ and $\nu$. The values of mean and effective speed are expressed in [$\mu m/min$].}
\label{V_Gaussian}
\end{table}

\noindent Moreover, we also include the cell tracking obtained in case $ii)$, i.e., with a non-uniform speed distribution, which allows to recover the reduced motility for higher values of the ECM density, but a random description of the fiber network.

\begin{figure}[!h]
\begin{center}
\includegraphics[width=0.3\textwidth]{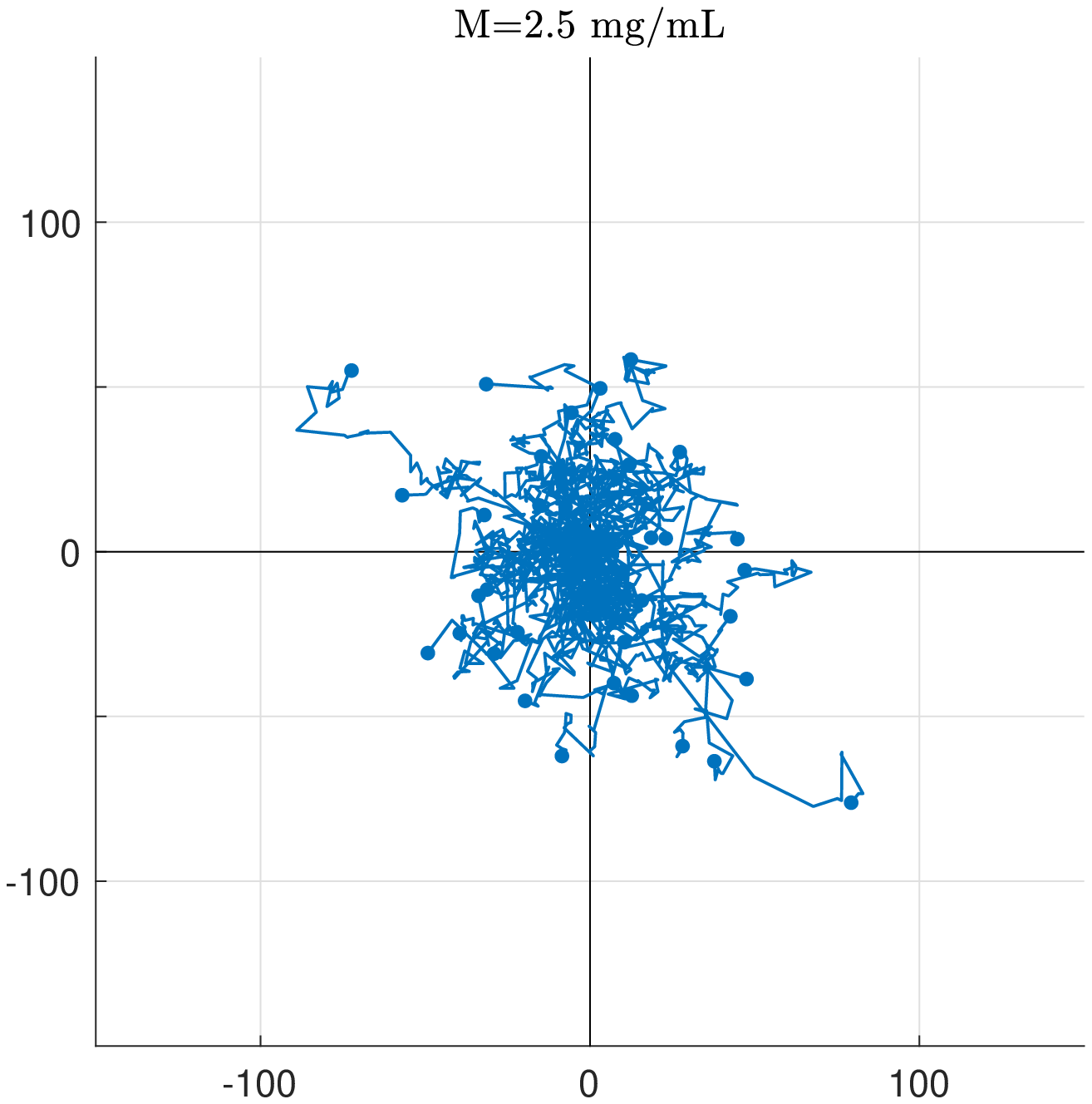}\hspace{0.3cm}
\includegraphics[width=0.3\textwidth]{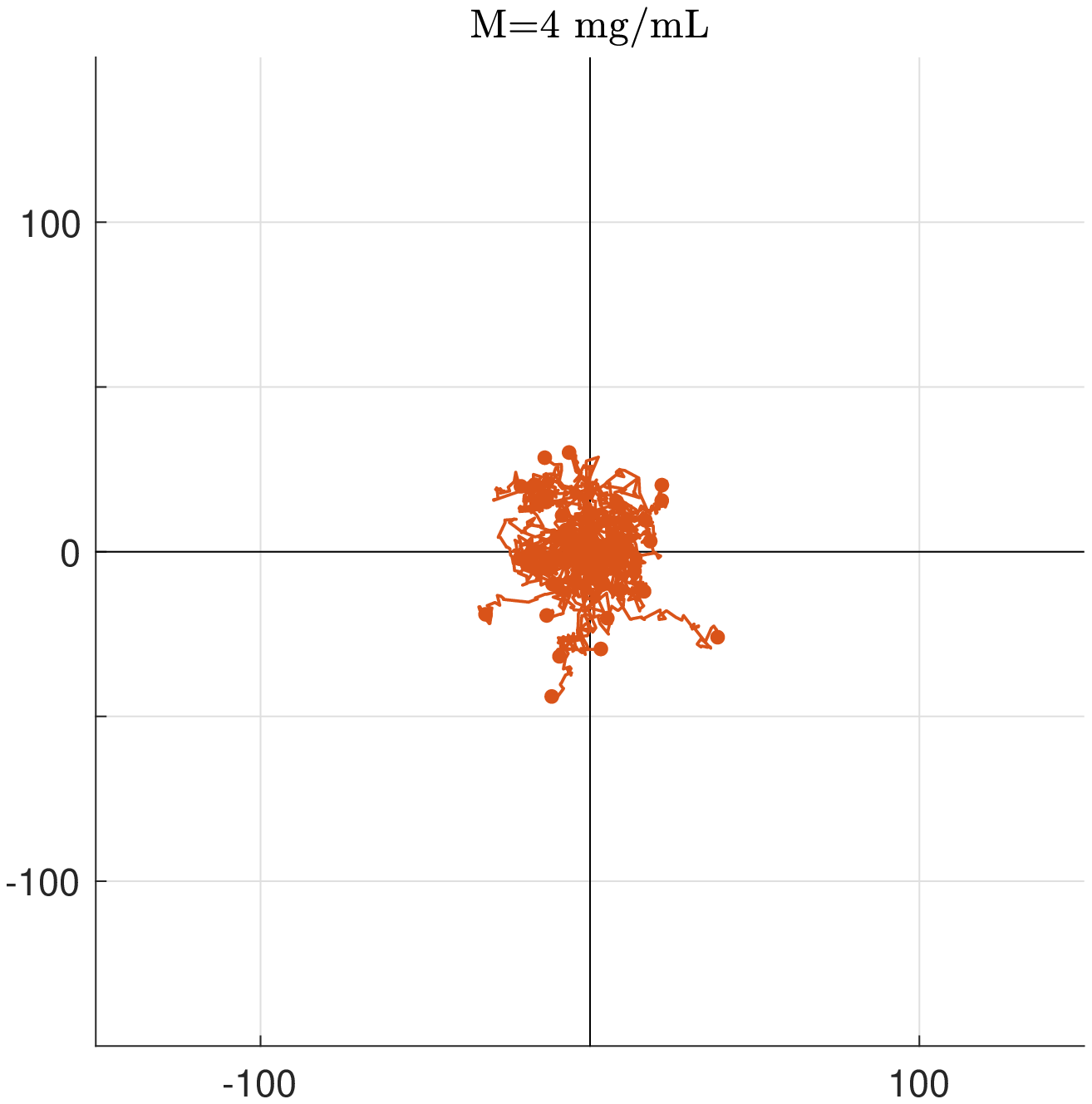}\hspace{0.3cm}
\includegraphics[width=0.3\textwidth]{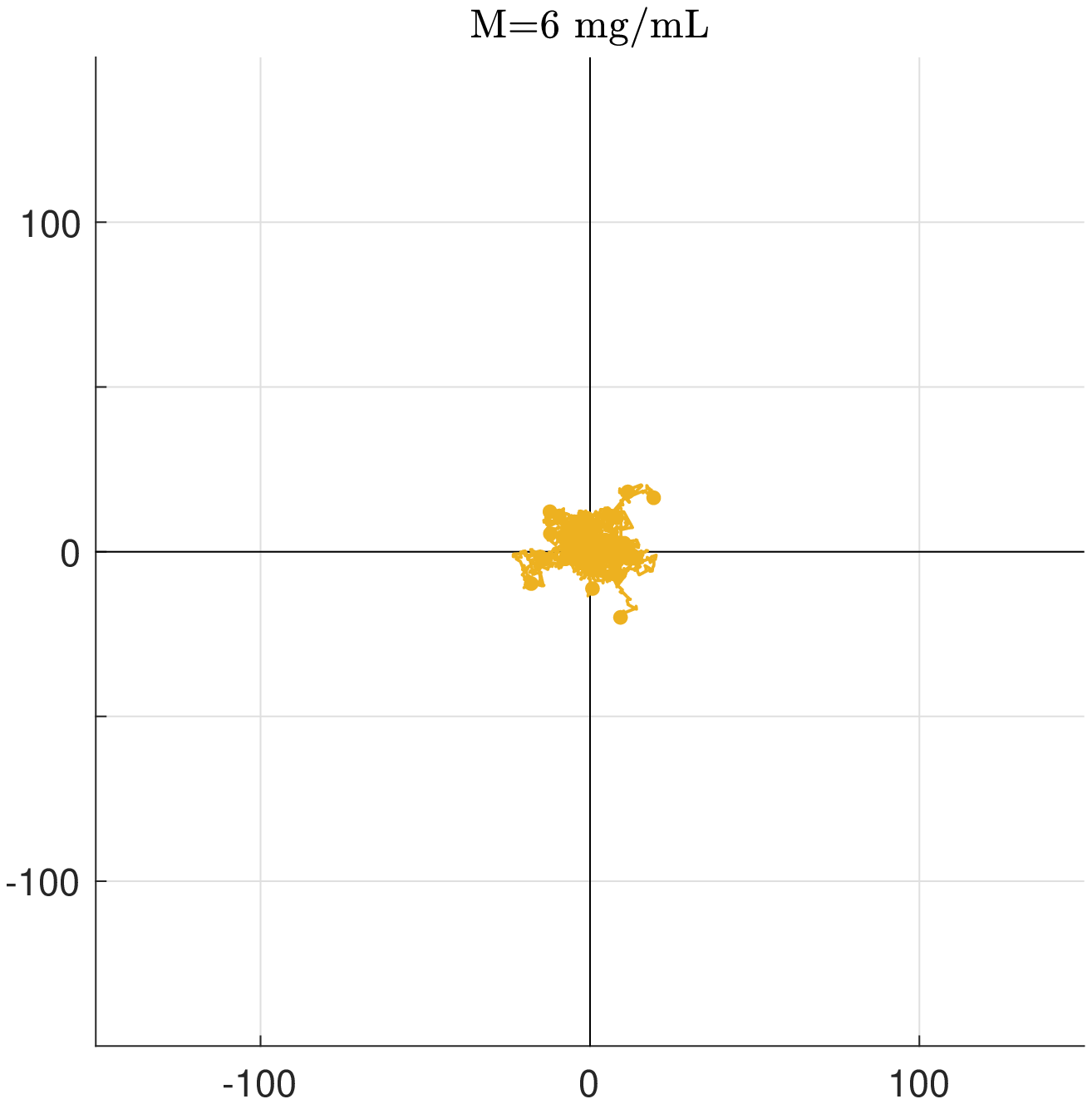}
\caption{{\bf Test 1.} Cell tracking in the domain $\Omega$ for the case $ii)$ of random fibers, assuming a truncated Gaussian distribution for the speed. Simulations are run for $24$h with $\Delta t=0.0167$ h.}
\label{fig_track_noalign}
\end{center}
\end{figure}

\section{The interplay of steric hindrance and fiber alignment in heterogeneous environments}\label{sec.num.4}
Here we propose an additional test to show the potential of the proposed approach to study cell behavior in a heterogeneous environment where the effects of fiber alignment on the migration are combined with the phenomenon of steric hindrance. We propose this test because non-locality is of the utmost importance when in presence of strongly heterogeneous or anisotropic environments. Such environments can be related both to the fiber structure \cite{han2016oriented}, as it may occur in presence of different TACS or at the tumor-stroma interface \cite{Carey2018,provenzano2006collagen}, and to the presence of interfaces between different ECM densities/porosities regions \cite{Sapudom2016AHM}.  

\begin{figure}[!h]
\centering
\includegraphics[width=0.5\textwidth]{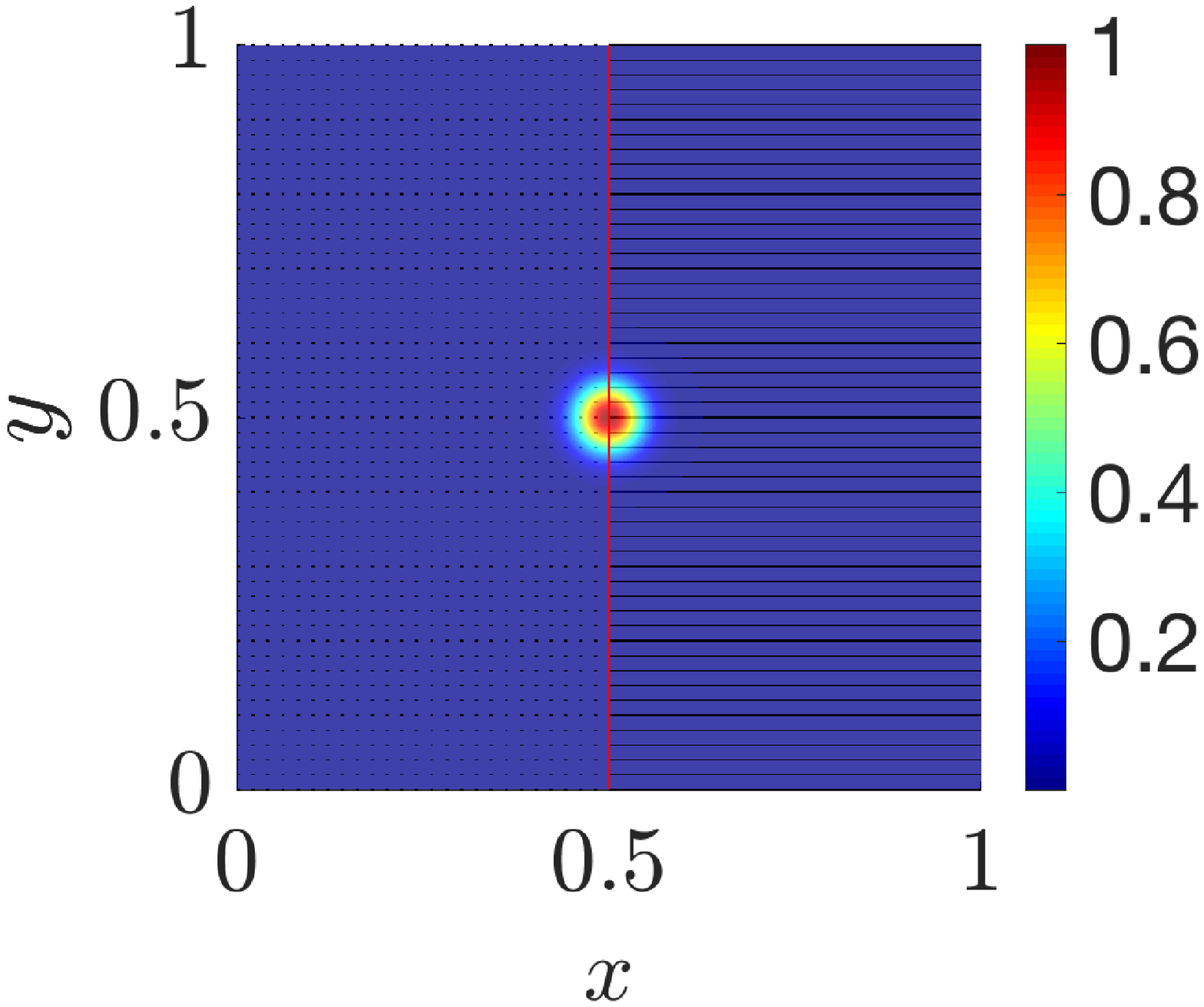}
\caption{\textbf{Test 4.} Initial distribution of the fibers in the two subregions $\Omega_1$ (on the left) and $\Omega_2$ (on the right) and initial Gaussian distribution of the cells, with mean $(x_0,y_0)=(50,50)$ and variance $\sigma^2=10^{-3}$. The vertical red line indicates the interface between the two subregions.}
\label{IC_test4} 
\end{figure}

\noindent Experimentally, the main motivation for this test comes from \cite{han2016oriented}, where the authors use a composite ECM made up of collagen of type $I$ and matrigel to determine the influence of the local collagen fiber orientation on the collective intravasation ability of breast cancer cells. They build a controllable and heterogeneous landscape with a homogeneous distribution of fibers inside the collagen and fibers vertically orientated near the interface between the collagen and the matrigel regions. They show how cells follow the local fiber alignment direction during the intravasation into rigid matrigel and how oriented fiber network could lead to a significantly enhanced infiltration potential. These results allow them to suggest the possibility of manipulating the ECM fiber structure orientation in the tumor microenvironment in order to alter and minimize the intravasation rate. The framework we propose allows us to actually analyze not only the impact of a heterogeneous landscape of fibers, but also to combine it with the effects of different collagen densities.\\
We consider the domain $\Omega=[0, 100]\times[0, 100]\, \mu m^2$ divided into two subregions $\Omega_1=[0, 50]\times[0 100]\,\mu m^2$ and $\Omega_2=(50 ,100] \times [0, 100]\, \mu m^2$. In $\Omega_1$ we consider an isotropic distribution of fibers, while in $\Omega_2$ we assumed that the fibers are oriented in the direction $\theta_q=0$ with alignment parameter $k=10$. In particular, we describe them using the bimodal von Mises distribution \eqref{bvm}  with the parameters specified above. Moreover, we consider two different values of matrix density $M_1=2.5\, mg/mL$ and $M_2=9.9\,mg/mL$ and the corresponding values for the mean speeds $\bar{v}_{M_1}=0.1696\, \mu m/min$ and $\bar{v}_{M_2}=0.01 \,\mu m/min$. We describe the speed distribution using the unimodal von Mises distribution \eqref{uvm}, where we set the value of the maximum cell velocity to $U=0.4\, \mu m/min$, while the concentration parameter is set to $k_\psi=100$. We consider an initial Gaussian distribution of cells centered in $(x_0,y_0)=(50,50)$ and with variance $\sigma^2=10^{-3}$ and we analyze cell behavior in two different scenario. In the first one, we assume that in $\Omega_1$ we have the matrix density $M_1$, while in $\Omega_2$ we have the matrix density $M_2$. In the second one, instead, we invert the values of the matrix density, setting $M_2$ in $\Omega_1$ and $M_1$ in $\Omega_2$. The initial configuration of fibers and cells used in this setting is shown in Fig. \ref{IC_test4}.

\begin{figure}[!ht]
\subfigure[t=1.875]{\includegraphics[width=0.32\textwidth]{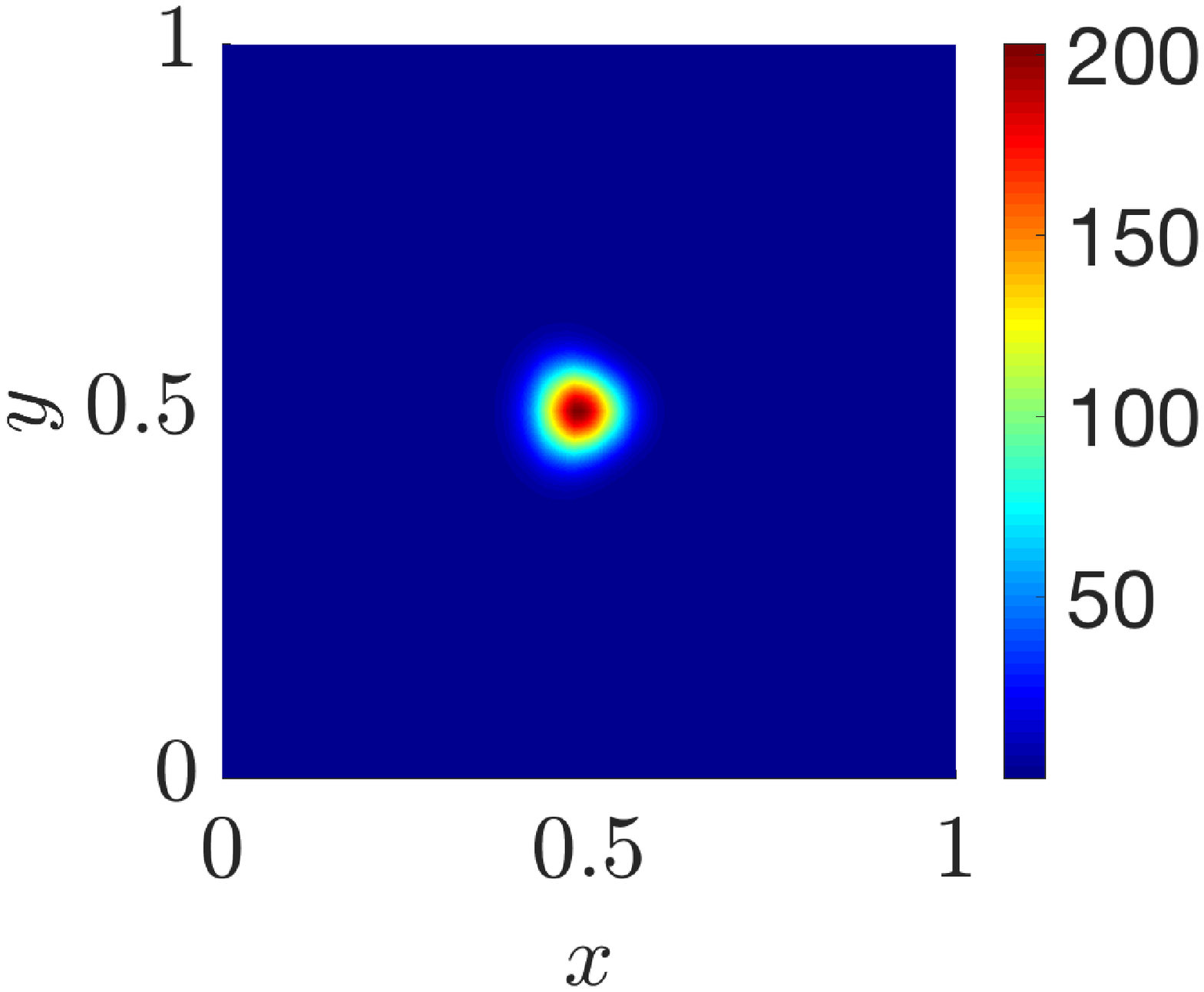}}
\subfigure[t=3.75]{\includegraphics[width=0.32\textwidth]{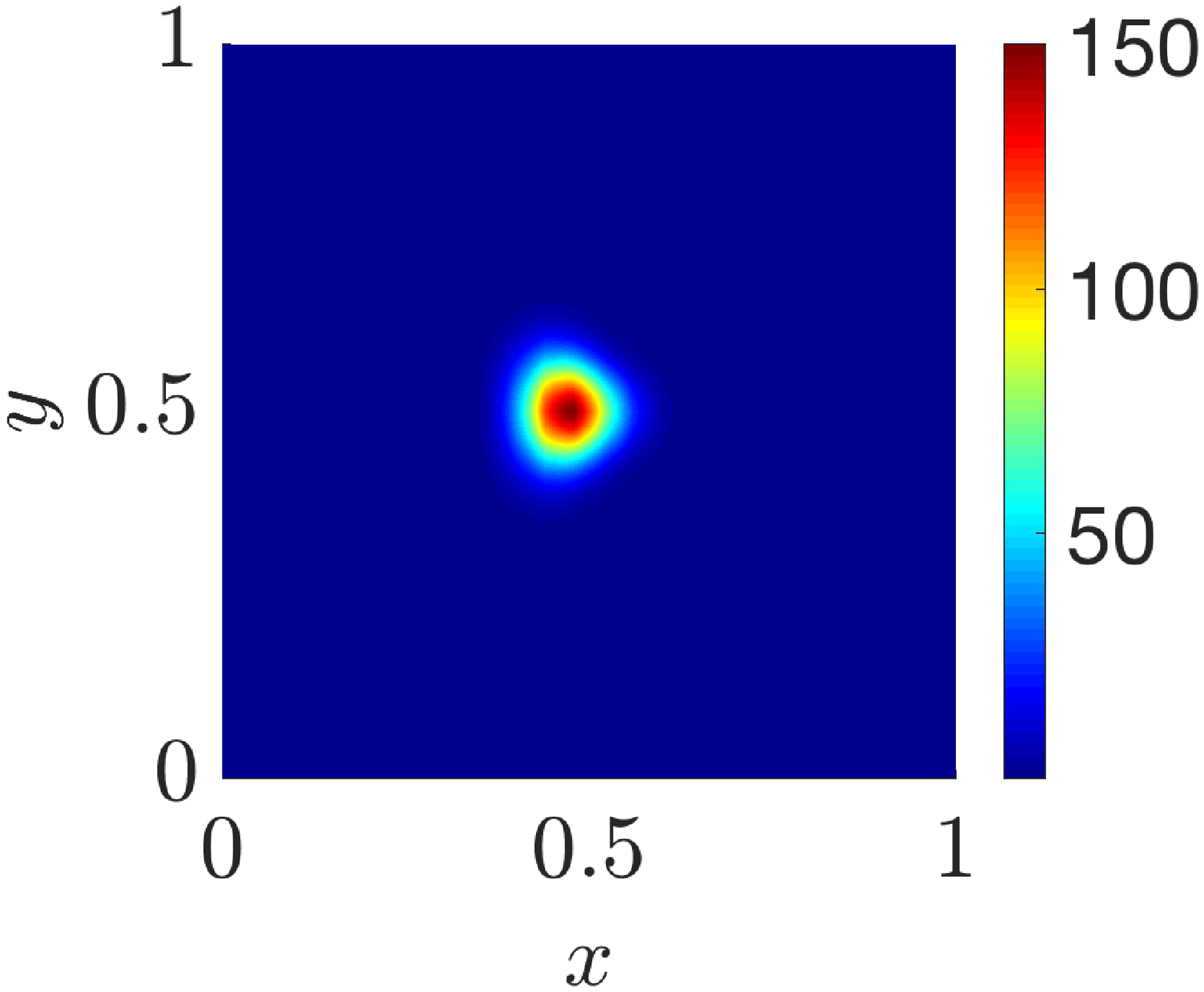}}
\subfigure[t=7.5]{\includegraphics[width=0.32\textwidth]{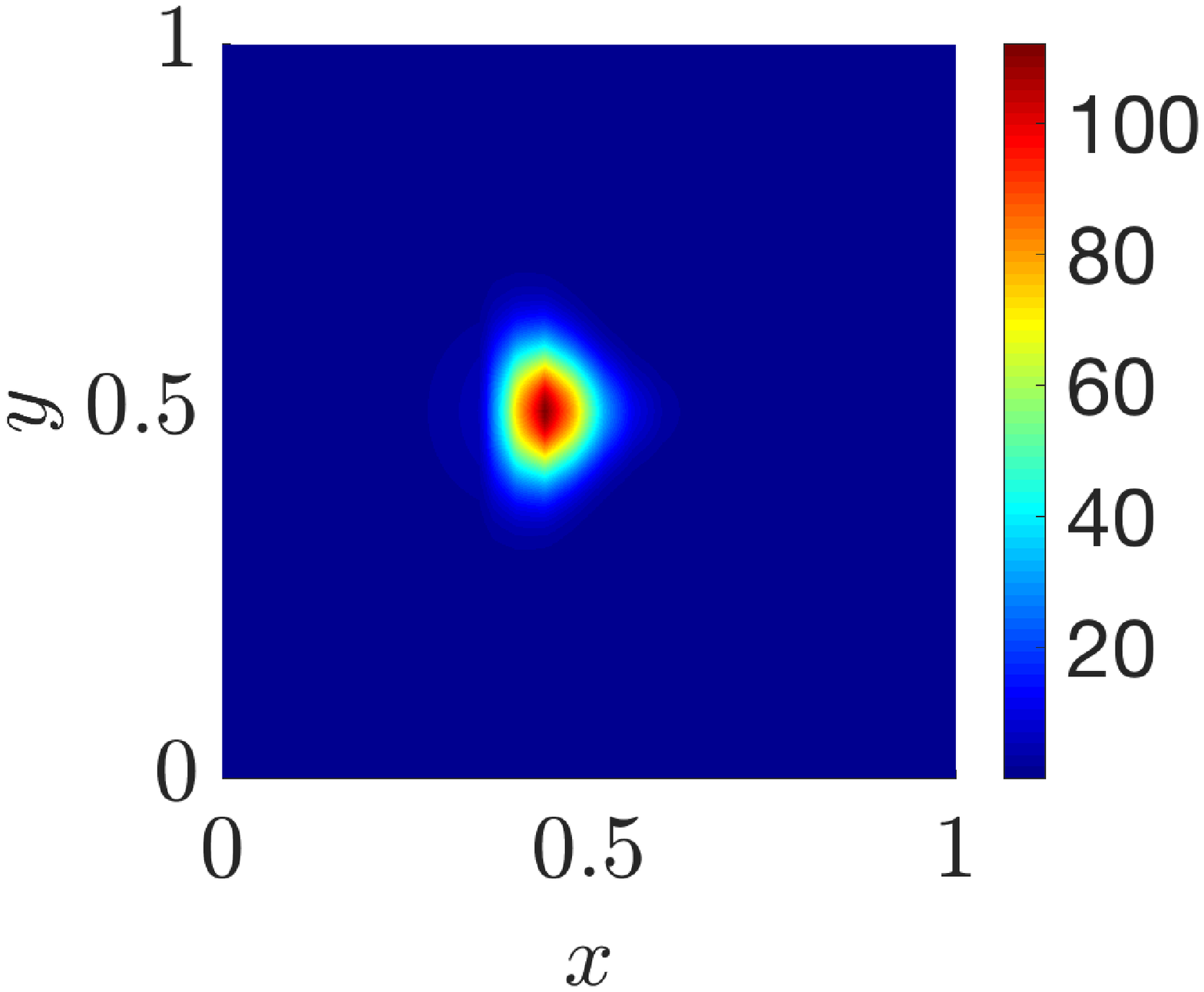}}\\
\subfigure[t=1.875]{\includegraphics[width=0.32\textwidth]{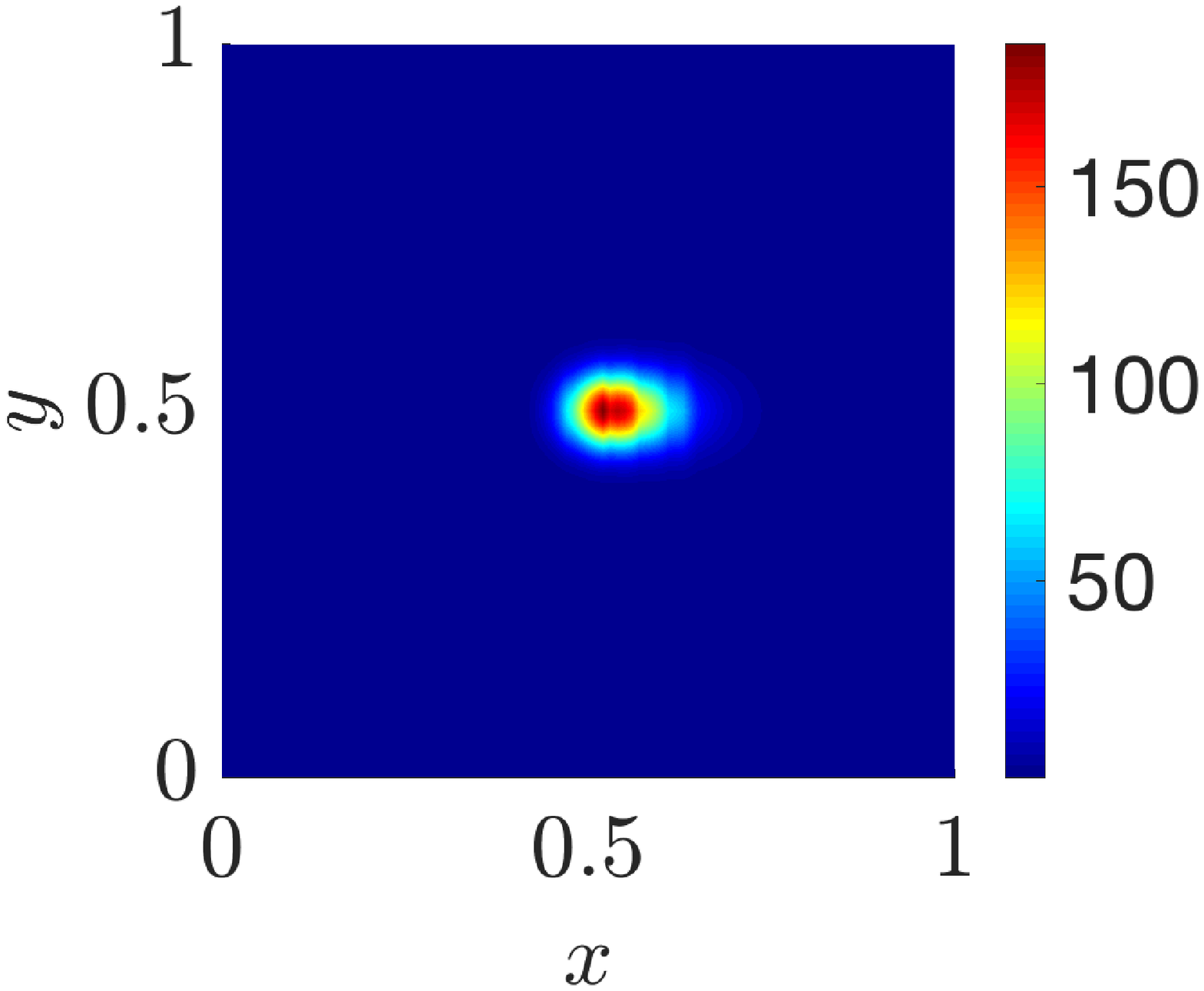}}
\subfigure[t=3.75]{\includegraphics[width=0.32\textwidth]{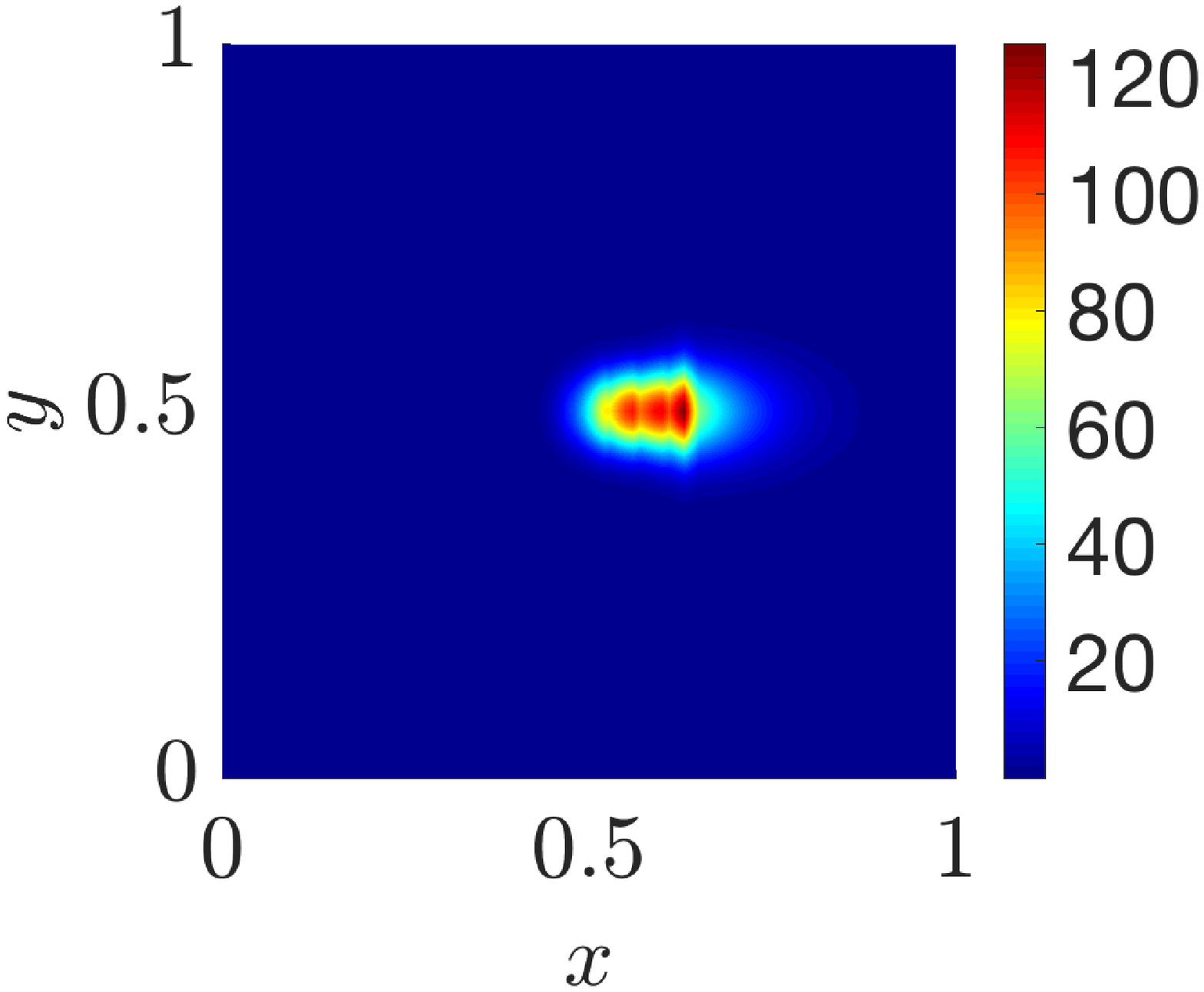}}
\subfigure[t=7.5]{\includegraphics[width=0.32\textwidth]{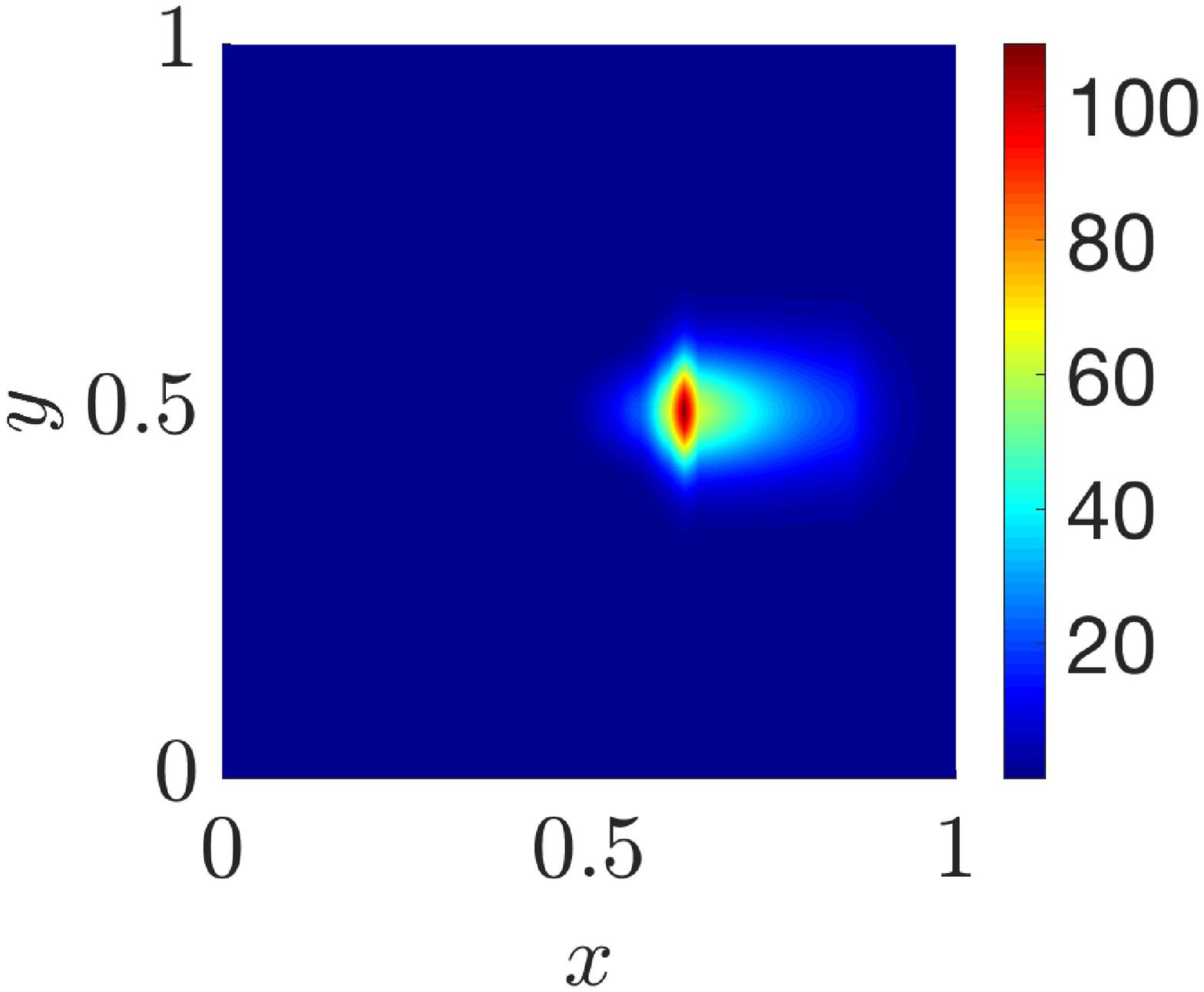}}
\caption{{\bf Test 4.} Qualitative evolution of the macroscopic cell density obtained from \eqref{eq:transp}-\eqref{eq:turning_operator}-\eqref{eq:trans1} with an initial Gaussian distribution of cell with mean $(x_0,y_0)=(50,50)$ and variance $\sigma^2=10^{-3}$. The sensing function $\gamma=\delta(R-\lambda)$. The first row refers to the case $M_1$ in $\Omega_1$ and $M_2$ in $\Omega_2$, while the second row refers to the case $M_2$ in $\Omega_1$ and $M_1$ in $\Omega_2$.}
\label{fig_IE_delta}
\end{figure}

\begin{figure}[!ht]
\subfigure{\includegraphics[width=0.48\textwidth]{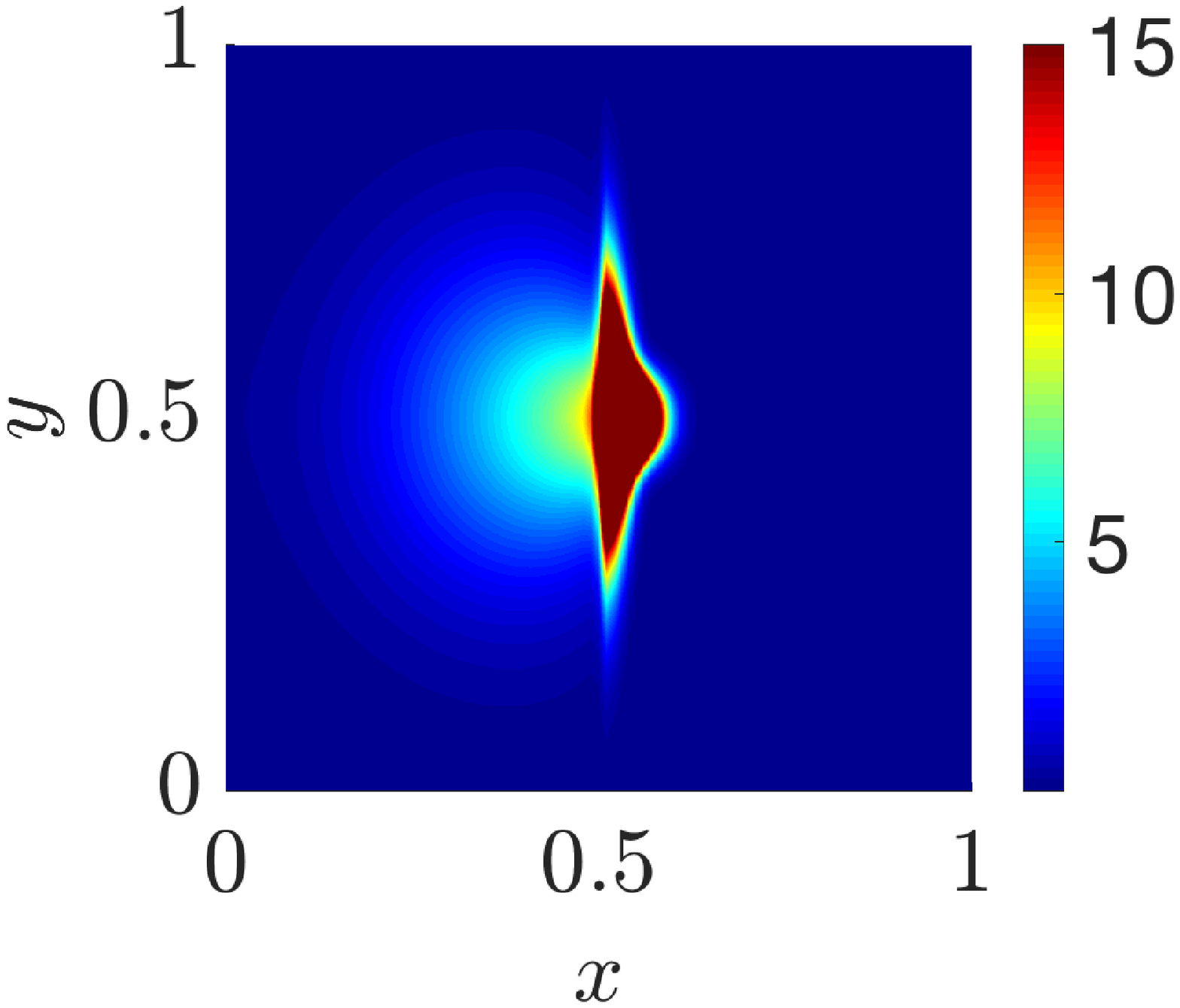}}
\subfigure{\includegraphics[width=0.48\textwidth]{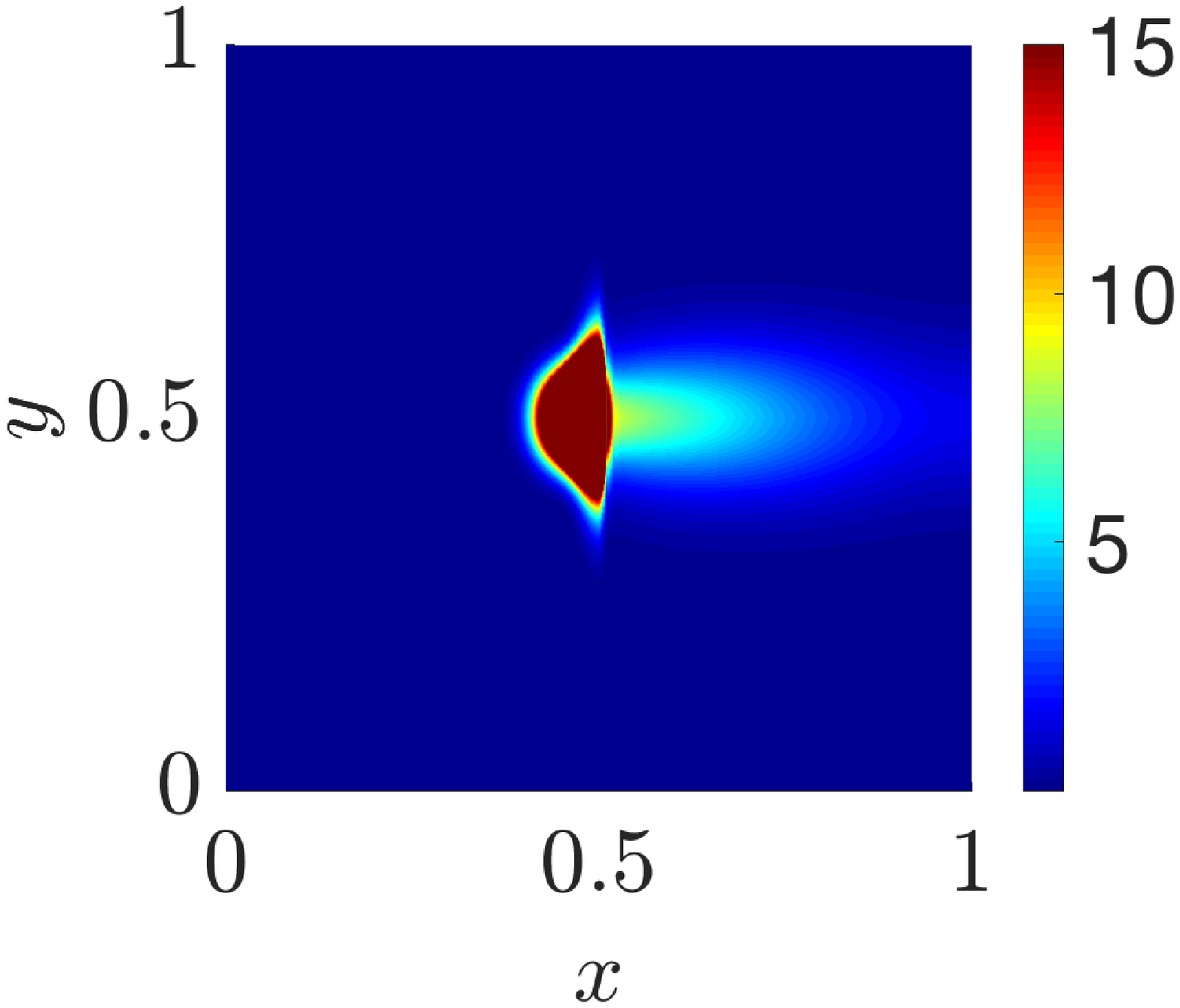}}
\caption{{\bf Test 4.} Evolution of the macroscopic cell density at $t=7.5$ obtained from \eqref{eq:transp}-\eqref{eq:turning_operator}-\eqref{eq:trans1} with an initial Gaussian distribution of cell with mean $(x_0,y_0)=(50,50)$ and variance $\sigma^2=10^{-3}$. The sensing function is $\gamma=\delta(\lambda-0)$. The left plot refers to the case $M_1$ in $\Omega_1$ and $M_2$ in $\Omega_2$, while the right plot refers to the case $M_2$ in $\Omega_1$ and $M_1$ in $\Omega_2$.}
\label{fig_IE_loc}
\end{figure}

\noindent Concerning the sensing function $\gamma$ in \eqref{eq:trans1}, we preliminary choose for this test a Delta function.
%(results for a different choice od the sensing function are provided in the Supplementary Fig. \ref{fig_IE_delta}). 
The qualitative evolution of the macroscopic cell density obtained by solving the kinetic model \eqref{eq:transp}-\eqref{eq:turning_operator}-\eqref{eq:trans1} in this setting are shown in Fig. \ref{fig_IE_delta}. The first row represent the case in which we set $M_1$ in $\Omega_1$ and $M_2$ in $\Omega_2$, while the second row the case $M_2$ in $\Omega_1$ and $M_1$ in $\Omega_2$.
In particular, in both scenarios we observe how in the region with collagen density $M_2$ (subregion $\Omega_2$ for the first row setting, while subregion $\Omega_1$ for the second row setting) the migration of the cells is highly reduced, while it is promoted toward the region where the collagen density is set to $M_1$. Moreover, we can observe that the way cells invade this region differs between the two experiments due to the underlying distribution of the fibers. In fact, the isotropic fiber distribution in $\Omega_1$ determines a more homogenous spreading of the cells, while the anisotropic fiber distribution in $\Omega_2$ determines a stronger cell alignment along the direction $\theta_q=0$. Furthermore, we notice the effect of non-locality in cell sensing of the microenvironment. In fact, in both cases, cells that are initially located in the region with the highest matrix density, but that are close enough (i.e., located at a distance lower than the sensing radius $R_M(t,\x,\hv)$ from the interface) to sense the environment in the more favorable region move towards it, instead of getting stuck due to the physical obstacle determined by the dense matrix. This is particularly clear if we compare these results with the evolution of the nonlocal kinetic model \eqref{eq:transp}-\eqref{eq:turning_operator}-\eqref{eq:trans1} with $\gamma=\delta(\lambda-0)$, shown in Fig. \ref{fig_IE_loc}, in which cells in the unfavorable region are not actually able to escape from it and their migration results highly limited.

\newpage

\bibliographystyle{siamplain}
\bibliography{references}